\documentclass[aps,prb,twocolumn,showpacs,superscriptaddress]{revtex4}
\usepackage{epsfig}
\usepackage{graphicx}
\usepackage{amsfonts}
\usepackage[figuresright]{rotating}
\usepackage{amssymb}
\usepackage{amsmath}
\usepackage{psfrag}

\def\avg#1{\langle#1\rangle}
\def\Re{\rm{Re}}

\def\be{\begin{equation}}
\def\ee{\end{equation}}
\def\bea{\begin{eqnarray}}
\def\eea{\end{eqnarray}}

\def\nn{\nonumber}
\def\pp{\parallel}
\def\tr{\mbox{tr}}
\def\Re{\mbox{Re}}

\begin{document}
\title{Unconventional symmetries of Fermi liquid and Cooper pairing properties
with electric and magnetic dipolar fermions}
\author{Yi Li}
\affiliation{Princeton Center for Theoretical Science,
Princeton University, Princeton, NJ 08544}
\author{Congjun Wu}
\affiliation{Department of Physics, University of California, San Diego,
CA 92093, USA}

\begin{abstract}
The rapid experimental progress of ultra-cold dipolar fermions opens up
a whole new opportunity to investigate novel many-body physics of fermions.
In this article, we review theoretical studies of the Fermi liquid theory and
Cooper pairing instabilities of both electric and magnetic dipolar
fermionic systems from the perspective of unconventional symmetries.
When the electric dipole moments are aligned by the external electric field,
their interactions exhibit the explicit $d_{r^2-3z^2}$ anisotropy.
The Fermi liquid properties, including the single-particle spectra,
thermodynamic susceptibilities, and collective excitations,
are all affected by this anisotropy.
The electric dipolar interaction provides a mechanism for
the unconventional spin triplet Cooper pairing, which is different
from the usual spin-fluctuation mechanism in solids and the superfluid $^3$He.
Furthermore, the competition between pairing instabilities in the
singlet and triplet channels gives rise to a novel time-reversal symmetry
breaking superfluid state.
Unlike electric dipole moments which are induced by electric fields
and unquantized, magnetic dipole moments are intrinsic proportional
to the hyperfine-spin operators with a Lande factor.
Its effects even manifest in unpolarized systems exhibiting an isotropic
but spin-orbit coupled nature.
The resultant spin-orbit coupled Fermi liquid theory supports a
collective sound mode exhibiting a topologically non-trivial spin
distribution over the Fermi surface.
It also leads to a novel $p$-wave spin triplet Cooper pairing state
whose spin and orbital angular momentum are entangled to the total
angular momentum $J=1$ dubbed the $J$-triplet pairing.
This $J$-triplet pairing phase is different from both the spin-orbit
coupled $^3$He-$B$ phase with $J=0$ and the spin-orbit
decoupled $^3$He-$A$ phase.
\end{abstract}

\pacs{Keywords: electric and magnetic dipolar interactions,
anisotropic Fermi liquid theory, SO coupled Fermi liquid
theory, $p$-wave triplet Cooper pairing, time-reversal symmetry
breaking
}
\maketitle

%*********************************************************************
%*********************************************************************
\section{Introduction}
\label{sect:intro}

Dipolar interactions have become a major research focus of ultra-cold
atomic and molecular physics.
For bosonic atoms with large magnetic dipolar moments (e.g. $^{52}$Cr),
their magnetic moments are aligned in the Bose-Einstein condensates
in which the anisotropy of the dipolar interaction is manifested.
\cite{lahaye2009,lahaye2008,menotti2008,menotti2008a,griesmaier2005,
trefzger2011,lian2012}.
On the other hand, the synthesis and cooling of
both fermions with electric and magnetic dipolar moments
give rises to an even more exciting opportunity to
explore novel many-body physics
\cite{ni2008,ospelkaus2010,ni2010,chotia2012,wuch2012,yan2013,hazzard2014,
zhuB2014,syzranov2014,wall2014,aikawa2014,aikawa2014a,aikawa2014b,
lu2010,lu2011,youn2010,burdick2014,baumann2014,lu2012}.
The quantum degeneracy of the fermionic dipolar molecules
of $^{40}$K$^{87}$Rb
has been realized \cite{ospelkaus2010,ni2008,ni2010}.
These molecules have been loaded into optical lattices
in which the loss rate is significantly suppressed \cite{zhuB2014,chotia2012}.
The chemically stable dipolar molecules of $^{23}$Na$^{40}$K have been
cooled down to nearly quantum degeneracy with a lifetime reported
as 100ms near the Feshbach resonance \cite{wuch2012}.
The quantum degeneracy of fermionic atoms with large magnetic dipole
moments has also been achieved for the systems of
$^{161}$Dy with 10$\mu_B$ \cite{lu2010,lu2010a,lu2012} and
$^{167}$Er with 7 $\mu_B$ \cite{aikawa2014,aikawa2014a,aikawa2014b},
which are characterized by the magnetic dipolar interaction.

%including the anisotropic Cooper pairing, Wigner crystallization,
%anisotropic distortion of the Fermi surfaces, nematic and stripe Fermi
%surface instabilities, collective modes of charge and spin excitations.

Electric and magnetic dipolar fermions exhibit novel many-body physics
that is not well-studied in usual solids.
One of the most prominent features of the electric dipolar interaction
is spatial anisotropy, which is markedly different from the
isotropic Coulomb interaction in solids.
In contrast, the magnetic dipolar interaction remains isotropic
in unpolarized systems.
More importantly, it exhibits the spin-orbit (SO) coupled feature,
{\it i.e.}, the magnetic dipolar interaction is invariant only
under the simultaneous rotation of both the orientations of magnetic
moments and their relative displacement vectors.
These features bring interesting consequences to the many-body
physics of dipolar fermions.

Rigorously speaking, so far there are still no permanent electric
dipole moments having been discovered yet at the levels of the
elementary particle, atom, and molecule.
For example, for a hetero-nuclear dipolar molecule, even though at
an instantaneous moment, it exhibits a dipole moment, while it
averages to zero in the molecular rotational eigenstates.
External electric fields are needed to polarize electric dipole
moments, which mixes rotational eigenstates with opposite parities.
However, the dipole moment of these mixed states is unquantized,
and, thus the electric dipole moment is a classic vector.
When two dipole moments are aligned, say, along the $z$-axis,
the interaction between them is spatially anisotropic,
which not only depends on the distance between
two dipoles, but also the direction of the relative displacement vector.
Nevertheless, this anisotropy exhibits an elegant form of the
spherical harmonics of the second Legendre polynomial,
{\it i.e.}, the $d_{r^2-3z^2}$-type anisotropy \cite{you1999,
baranov2002,baranov2012}.
This elegant anisotropy greatly simplifies the theoretical
study of the novel many-body physics with the electric dipolar
interaction.

The electric dipolar interaction results in an anisotropic Fermi
liquid state, which exhibits different single-particle and collective
properties from those of the standard isotropic Fermi liquid theory
\cite{miyakawa2008,sogo2009,fregoso2009a,chan2010,sun2010,
ronen2010,lin2010,liu2011,liDasSarma2010,kestner2010,
baranov2008a,rodriguez2014}.
The shape of the Fermi surface exhibits anisotropic distortions
\cite{miyakawa2008,sogo2009,fregoso2009,chan2010,kestner2010}.
The anisotropic dipolar interaction mixes different partial-wave
channels, and thus the usual Landau interaction parameters in
the isotropic case should be generalized into the Landau interaction matrix
with a tri-diagonal structure, which renormalizes thermodynamic
susceptibilities \cite{fregoso2009a,chan2010}.
The dispersion of the collective zero sound mode is also anisotropic:
the zero sound mode can only propagate in a certain range of the solid angle
direction, and its sound velocity is maximal if the propagation
direction is along the north or south poles
\cite{chan2010,ronen2010}.

The anisotropy of the electric dipolar interaction also results in
unconventional Cooper pairing symmetries
\cite{you1999,baranov2002,baranov2004,
bruun2008,levinsen2011,potter2010,lutchyn2010,
zhao2010,wu2010,shi2009,qi2013,shi2013,liu2014}.
The electric dipolar interaction is neither purely attractive nor
purely repulsive.
The partial-wave analysis shows that the most attractive pairing
channel is $p_z$-like, which naturally gives rise to a new
mechanism to unconventional pairing symmetry.
Consequently, for the single component case, the pairing symmetry
is mostly of $p_z$-like slightly hybridized with even higher
odd partial wave components \cite{baranov2002,baranov2004,you1999,bruun2008}.
The pairing structure of the two-component dipolar fermions is
even more interesting, which allows both the $s+d$-wave channel
singlet and the $p_z$-wave triplet pairings
\cite{wu2010,shi2010,qi2013,shi2013}.
The dipolar interaction induced triplet pairing
is to first order in interaction strength.
In comparison, the spin fluctuation mechanism in solid state systems
(e.g. $^3$He and Sr$_2$RuO$_4$) is a higher order effect of
interactions \cite{leggett1975,mackenzie2003}.
The singlet and triplet pairing symmetries can coexist
in two-component electric dipolar fermion systems.
Only when their relative phase angle is $\pm \frac{\pi}{2}$,
the resultant pairing is unitary \cite{wu2010}.
This gives rise to a novel and very general mechanism to a spontaneous
time-reversal (TR) symmetry breaking pairing state.

Next we discuss the novel feature of the magnetic dipolar fermions
\cite{fujita1987,fregoso2009,fregoso2010,li2012,sogo2012,li2012a,bhongale2013,
tohyama2013,ashrafi2013}.
The magnetic dipolar interaction is very complicated to handle in
classic systems, which leads to a variety of rich patterns in
real space.
In comparison, for the quantum degenerate Fermi systems, the existence
of Fermi surfaces constraints the low energy degrees of freedom only
around the Fermi surface.
This feature greatly simplifies the theoretical analysis, and the
exotic physics with non-trivial spin texture patterns lies in
momentum space instead of real space.

Typically speaking, the interaction energy scale of magnetic
dipolar fermions is much smaller than that of the electric dipolar case.
Nevertheless, conceptually they are still very interesting.
Unlike the electric dipolar moment, the magnetic moment is proportional
to the hyperfine-spin with the Lande factor, and thus its components
are non-commutative quantum-mechanical operators
\cite{li2012,li2012a}.
Magnetic dipole moments are permanent in the sense that they
do not need to be induced by external magnetic fields.
In the absence of external fields, the unpolarized magnetic dipolar
systems are in fact isotropic.
Neither spin nor orbital angular momentum is conserved; %,
nevertheless, the total angular momentum
remains conserved by the dipolar interaction.
Thus the magnetic dipolar interaction exhibits the essential
feature of the SO coupling physics.
Very recently, using electric dipolar moments to generate effective SO
coupled interactions similar to that in the magnetic dipolar systems
is also proposed in
Ref. [\onlinecite{wall2014}] by properly coupling microwaves to
molecular rotation eigenstates.

The ordinary SO coupling in solids is a single-particle effect originating
from the relativistic physics.
In contrast, in magnetic dipolar fermion systems
\cite{fregoso2010,sogo2012,li2012a},
the Fermi surfaces remain spherical
without splitting in the absence of the external magnetic fields.
Nevertheless,  this SO coupling appears at the interaction level,
including the SO coupled Fermi surface Pomeranchuk instabilities
\cite{sogo2012,li2012a,fregoso2010}, and
topological zero-sound wave modes exhibiting an oscillating
spin distribution of the hedgehog-type configuration over the Fermi surface
\cite{li2012a}.

The magnetic dipolar interaction also induces novel Cooper pairing
structures exhibiting the SO coupled nature \cite{li2012,fishman1987}.
Even in the simplest case of $F=\frac{1}{2}$, the magnetic dipolar
interaction provides a novel and robust mechanism for the $p$-wave ($L=1$)
spin triplet ($S=1$) Cooper pairing which arises from the attractive
channel of the magnetic dipolar interaction.
It turns out that its pairing symmetry structure is markedly
different from that in the celebrated $p$-wave pairing system of
$^3$He: the orbital angular momenta $L$ and spin $S$ of
Cooper pairs are entangled into the channel of the
total angular momentum  $J = 1$,
dubbed as the $J$-triplet pairing.
In comparison, the $^3$He-$B$ phase is isotropic in which
$J=0$, while the $A$-phase is anisotropic in which $J$
is not well-defined \cite{leggett1975}.

In this article, we review the recent progress of the novel many-body
physics with dipolar fermions, such as the Fermi liquid properties and
Cooper pairing structures, focusing on unconventional symmetries.
In Sect. \ref{sect:fourier}, we review the anisotropy of the
electric dipolar interaction, and the SO structure of the magnetic
dipolar interactions, respectively,
from the viewpoint of their Fourier components.
In  Sect. \ref{sect:el_dp_FL}, the anisotropic Fermi liquid theory
of the electric dipolar fermions is reviewed. %, a
And the SO coupled
Fermi liquid theory of the magnetic dipolar fermion systems
is reviewed in Sect. \ref{sect:mg_dp_FL}.
The $p_z$-wave Cooper pairing in the single and two-component
electric dipolar systems and the TR reversal symmetry breaking
effect are reviewed in Sect. \ref{sect:el_dp_pair}.
The SO coupled Cooper pairing with the $J$-triplet structure
in the magnetic dipolar fermion systems is reviewed
in Sect. \ref{sect:mg_dp_pair}.
Conclusions and outlooks are presented in Sect. \ref{sect:conclusion}.

Due to limit of space and also the view point from the unconventional symmetry,
we are not able to cover many important research
directions of dipolar atoms and molecules in this review.
For example, the progress on topics of strong correlation physics with
dipolar fermions \cite{wang2010,bhongale2012,han2010}, the Feshbach
resonance %resonaces
with dipolar fermions \cite{qi2013,shi2013},
the synthetic gauge field with dipolar fermions \cite{deng2012,cui2013},
and the engineering of exotic and topological many-body states
\cite{yao2012,yao2013,kestner2011}.
Some of these progresses have been excellently reviewed in
Ref. [\onlinecite{baranov2012,wall2014}].
The properties of dipolar boson condensations are not covered here either,
and there are already many important reviews on this topic
\cite{baranov2012,baranov2008,lahaye2009,menotti2008,
menotti2008,yisu2008}.

%%%%%%%%%%%%%%%%%%%%%%%%%%%%%%%%%%%%%%%%%%%%%%%%%%%%%%%%%%%%%%%%%%
\section{Fourier transform of dipolar interactions}
\label{sect:fourier}

In this section, we review the Fourier transformations of both the
electric and magnetic dipolar interactions in Sect. \ref{sect:four_el}
and Sect. \ref{sect:four_mag}, respectively.
The anisotropy of the electric dipolar interaction
and the SO coupled feature of the magnetic dipolar interaction
also manifest in their momentum space structure.
These Fourier transforms are important for later analysis
of many-body physics.

%-----------------------------------------------
\subsection{Electric dipolar interaction}
\label{sect:four_el}

Without loss of generality, we assume that all the electric dipoles
are aligned by the
external electric field $\vec E$ along the $z$-direction, then
the dipolar interaction between two dipole moments is
\cite{you1999,baranov2002}
\bea
V_d(\vec r_{12})
&=& -{2 d^2 \over {|r_{12}|^3}} P_2 (\cos \theta _{12}),
\label{eq:interaction_3D}
\eea
where $d$ is the magnitude of the electric dipole moment;
$\vec r_{12}=\vec r_1-\vec r_2$ is the displacement vector between two dipoles;
$\theta_{12}$ is the polar angle of $\vec r_{12}$;
$P_2(\cos\theta_{12})$ is the standard second Legendre polynomial as
$$P_2 (\cos \theta _{12})=\frac{1}{2} (3 \cos^2 \theta_{12}-1).$$
The zeros of the second Legendre polynomial lie around
the latitudes of $\theta_0$ and $\pi-\theta_0$ with
\bea
\theta_0=\cos^{-1}\frac{1}{\sqrt 3}\approx 55^\circ.
\label{eq:theta0}
\eea
Within $\theta_0<\theta_{12}<\pi-\theta_0$, the dipolar interaction
is repulsive, and otherwise, it is attractive.
The spatial average of the dipolar interaction in 3D is zero.

For later convenience, we introduce the Fourier transform of
the dipolar interaction Eq. \ref{eq:interaction_3D},
\bea
V_d(\vec q)=\int d^3 \vec r e^{-i \vec q \cdot \vec r} V_d(\vec r).
\label{eq:fourier}
\eea
A lot of information can be obtained solely based on symmetry
and dimensional analysis.
First, $e^{i\vec q \cdot \vec r}$ is invariant under spatial
rotations, thus $V_d(\vec q)$ transforms the same as $V_d(\vec r)$
under spatial rotations.
It should exhibit the same symmetry factor of the spherical harmonics.
Second, since $V_d(\vec r)$ decays with a cubic law, $V_d(\vec q)$
should be dimensionless. %in terms of the length unit.

If $V_d(\vec r_{12})$ were isotropic,
$V_d(\vec q)$ would logarithmically depends on $q$.
However, a more detailed calculation shows that actually it does not
depend on the magnitude of $q$.
Let us introduce a short distance cutoff $\epsilon$ that
the dipolar interaction Eq. \ref{eq:interaction_3D} is only
valid for $r\ge \epsilon$, and a long distance
cutoff $R$ as the radius of the system.
A detailed calculation shows that \cite{chan2010}
\bea V_{d}(\vec
q)&=&8\pi d^2 \left\{\frac{j_1(q\epsilon)}{q\epsilon} -\frac{j_1(q
R)}
{q R}\right\} P_2(\cos\theta_{\vec q})
\eea
where $j_1(x)$ is the first order spherical Bessel function
with the asymptotic behavior as
\bea
j_1(x)& \rightarrow&
\Big\{
\begin{array}{c}
\frac{x}{3}, \ \ \, \ \ \, \ \ \, \ \ \, \ \ \, \ \ \, \mbox{as} ~~~
x\rightarrow 0; \\
\frac{1}{x} \sin
( x- \frac{\pi}{2}), ~~~\mbox{as}~~~~ x\rightarrow \infty.
\end{array}
\eea
After taking the limits of  $q\epsilon\rightarrow 0$ and
$q R\rightarrow +\infty$, we arrive at
\bea V_d(\vec
q) = \frac{8\pi d^2}{3} P_2(\cos \theta_{\vec q}).
\label{eq:interaction_3Dk}
\eea
At $\vec q=0$, $V_d(\vec q=0)$ is defined as 0 based on the fact that
the angular average of the 3D dipolar interaction vanishes, thus $V_d$
is singular as $\vec q\rightarrow 0$.
Even in the case that $R$ is large but finite, the smallest nonzero
value of $qR$ is at the order of $O(1)$.
Thus,  $V_d (\vec q)$ remains non-analytic as $\vec q \rightarrow 0$.

\begin{figure}[tb]
\centering\epsfig{file=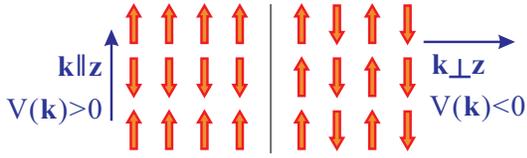,width=0.8\linewidth,angle=0}
\caption{The Fourier components of the dipolar interaction
$V_d(\vec k)$.
The left-hand-side is for $\vec k\parallel \hat z$, and the
right-hand-side is for $\vec k\perp \hat z$.
}
\label{fig:dip-fr}
\end{figure}

An interesting feature of the above Fourier transform Eq.
\ref{eq:interaction_3Dk} is that the anisotropy in momentum
space is opposite to that in real space: it is most
negative when $\vec q$ lies in the equatorial plane, and most
positive when $\vec q$ points to the north and south poles.
An intuitive picture is explained in Fig. \ref{fig:dip-fr}.
Consider a spatial distribution of the dipole density $\rho(r)$,
then the classic interaction energy is
\bea
\int \int dr_1 dr_2 \rho(\vec r_1) \rho(\vec r_2)
V_d(\vec r_1-\vec r_2)
&=&\frac{1}{V_0}\sum_{\vec q} |\rho(\vec q)|^2 \nn \\
&\times& V_d(\vec q),
\eea
where $V_0$ is the system volume.
If the wave vector $\vec q$ is along the $z$-axis, then the dipole
density oscillates along the dipole orientation, thus
the interaction energy is repulsive.
On the other hand, if $\vec q$ lies in the equatorial plane,
the dipole density oscillates perpendicular to the dipole
orientation, and thus the interaction energy is attractive.

%----------------------------------------------------------------
%---------------------------------------------------------------

%--------------------------------------------------------
\subsection{Magnetic Dipolar Interaction}
\label{sect:four_mag}

Now let us consider the magnetic dipolar interaction \cite{lu2010,
lu2010a,lu2012,aikawa2014,aikawa2014a,aikawa2014b}.
Different from the electric dipole moment, the magnetic one
originates from contributions of several different angular
momentum operators.
The total magnetic moment is not conserved, and thus its component
perpendicular to the total spin averages to zero.
For the low energy physics below the coupling energy among different
angular momenta, the magnetic moment can be approximated as
just the component parallel to the spin direction, and thus
the effective magnetic moment is proportional to the  hyperfine
spin operator up to a Lande factor, and thus is a quantum mechanical
operator.
Due to the large difference of energy scales between the fine
and hyperfine structure couplings, the effective atomic magnetic moment
below the hyperfine energy scale can be calculated through the
following two steps.
The first step is the Lande factor for the electron magnetic
moment respect to total angular momentum of electron defined as
$\vec \mu_e= g_J \mu_B \vec J$,
where $\mu_B$ is the Bohr magneton; $\vec J=\vec L +\vec S$ is
the sum of electron orbital angular momentum $\vec L$ and
spin $\vec S$; and the value of $g_J$ is determined as
\bea
g_J=\frac{g_L+g_s}{2} +\frac{g_L-g_s}{2} \frac{L(L+1)-S(S+1)}{
J(J+1)}.
\eea
Further considering the hyperfine coupling, the total magnetic
momentum is defined $\vec \mu= \mu_B( g_J\vec J+ g_I \vec I)
=g_F \vec F$ where $g_I$ is proportional to the nuclear gyromagnetic ratio
and is thus tiny, and $\vec F$ is the hyperfine spin.
The Lande factor $g_F$ can be similarly calculated as
\bea
g_F&=&\frac{g_J+g_I}{2}+\frac{g_J-g_I}{2}\frac{J(J+1)-I(I+1)}{F(F+1)}
\nn \\
&\approx& \frac{g_J}{2}\Big(1+\frac{J(J+1)-I(I+1)}{F(F+1)}\Big).
\eea

The magnetic dipolar interaction between two spin-$F$ atoms located
at $\vec r_{1}$ and $\vec r_2$ is
\bea
V_{\alpha \beta; \beta^{\prime}\alpha^{\prime} }(\vec r)
&=&\frac{g^2_F\mu_B^{2}}{r^{3}}
\Big[ \vec F_{\alpha \alpha^{\prime}}\cdot \vec F_{\beta \beta^{\prime}}
-
3 ( \vec F_{\alpha \alpha^{\prime}}\cdot \hat{r})
  ( \vec F_{\beta \beta^{\prime}}\cdot \hat{r}) \Big],\nn \\
\label{eq:mgdpr_r}
\eea
where $\vec r=\vec r_1 - \vec r_2$ and $\hat r=\vec r /r$ is
the unit vector along $\vec r$.
Similarly to the case of the electric dipolar interaction,
the Fourier transform of Eq. \ref{eq:mgdpr_r} possesses
the same symmetry structure as that in real space
\cite{fregoso2009,fregoso2010}
\bea
V_{\alpha \beta; \beta^{\prime}\alpha^{\prime} }(\vec q) &=&\frac{4 \pi g^2_F\mu_B^{2}}{3}
\Big[ 3( \vec {F}_{\alpha \alpha^{\prime}}\cdot \hat{q})
( \vec {F}_{\beta \beta^{\prime}}\cdot \hat{q})\nn \\
&-&\vec{F}_{\alpha \alpha^{\prime}}\cdot \vec{F}_{\beta \beta^{\prime}}
\Big].
\label{eq:four_mgdp}
\eea
Again, it only depends on the direction
of the momentum transfer but not on its magnitude, and
it is also singular as $\vec q \rightarrow 0$.
If $\vec q$ is exactly zero, $V_{\alpha\beta;\beta^\prime \alpha^\prime}(\vec q=0)=0$.

%The second quantization form of the magnetic dipolar interaction is
%expressed as
%\bea
%H_{md}&=&\sum_{\vec k,\alpha} [\epsilon(\vec k) -\mu]
%c^\dagger_\alpha(\vec k) c_\alpha(\vec k)
%+\frac{1}{2V_0}\sum_{\vec k,\vec k^\prime,\vec q}
%V_{\alpha\beta;\beta^\prime\alpha^\prime}(\vec q)
%\nn \\
%&\times& \psi^\dagger_\alpha(\vec k+\vec q)
%\psi^\dagger_\beta(\vec k^\prime)
%\psi_{\beta^\prime}(\vec k^\prime+\vec q) \psi_{\alpha^\prime}(\vec k).
%\eea
%Similarly, the dimensionless interaction parameter can be
%defined accordingly as  $\lambda_{md}= N_F \mu_B^2g_F^2$.

In the current experiment systems of magnetic dipolar atoms,
the atomic spin is very large.
For example, for  $^{161}$Dy, its atomic spin reaches $F=\frac{21}{2}$,
and thus an accurate theoretical
description of many-body physics of the magnetic dipolar interactions
of such a large spin system would be quite challenging
\cite{lu2010,lu2012}.
Nevertheless, as a theoretical starting point, we can use
the case of $F=\frac{1}{2}$ as a prototype model which exhibits
nearly all the qualitative features of the magnetic dipolar
interactions \cite{fregoso2009,li2012}.

%%%%%%%%%%%%%%%%%%%%%%%%%%%%%%%%%%%%%%%%%%%%%%%%%%%%%%%%%%%
%%%%%%%%%%%%%%%%%%%%%%%%%%%%%%%%%%%%%%%%%%%%%%%%%%%%%%%%%%%
%%%%%%%%%%%%%%%%%%%%%%%%%%%%%%%%%%%%%%%%%%%%%%%%%%%%%%%%%%%
\section{Anisotropic Fermi liquid theory of electric dipolar
fermions}
\label{sect:el_dp_FL}

In this section, we will review the new ingredients of the Fermi liquid
theory brought by the anisotropic electric dipolar interaction
\cite{miyakawa2008,sogo2009,fregoso2009a,chan2010,sun2010,
ronen2010,lin2010,liu2011,liDasSarma2010,baranov2008a,rodriguez2014},
including the single-particle properties such as Fermi surface distortions,
and two-body properties including thermodynamic properties
and collective modes.

A general overview of the Landau-Fermi liquid theory is presented
in Sect. \ref{sect:FL}.
In Sect. \ref{sect:fermisurface}, we review the dipolar interaction
induced Fermi surface distortions.
The Landau interaction matrix is presented in Sect.
\ref{sect:landau_matrix}, and its renormalization on thermodynamic
properties including Pomeranchuk instabilities are
review in Sect. \ref{sect:thermo}.
The anisotropic collective excitations are reviewed in
Sect. \ref{sect:zero_sound}.

%-----------------------------------------------------------
\subsection{A quick overview of the Fermi liquid theory}
\label{sect:FL}

One of the most important paradigms of the interacting fermion systems
is the Landau Fermi liquid theory \cite{landau1957,landau1959,negele1988}.
The Fermi liquid ground state can be viewed as an adiabatic evolution
from the non-interacting Fermi gas by gradually turning on interactions.
Although the ideal Fermi distribution function could be significantly
distorted, its discontinuity remains which still defines a Fermi surface
enclosing a volume in momentum space proportional to the total fermion
number, as stated by the Luttinger theorem.
Nevertheless the shape of the Fermi surface can be modified by
interactions.
The low energy excitations become the long-lived quasi-particles
around the Fermi surface, whose life-time is inversely proportional
to the square of its energy due to the limited phase space
for low energy scattering processes.
The overlap between the quasi-particle state and the bare fermion
state defines the wavefunction renormalization factor $Z$, which
is suppressed from the non-interacting value of 1 but remains
finite.
$Z$ is also the quasiparticle weight determining the discontinuity
of the fermion occupation number at the Fermi surface.

The interactions among quasi-particles are captured by
the phenomenological Landau interaction function, which describes
the forward scattering processes of quasi-particles.
The Landau interaction function can be decomposed into a set of
Landau parameters $F_l$ in which $l$ denotes the partial wave channels.
The physical observables, such as compressibility,
specific heat, and magnetic susceptibility, compared with their values
in free Fermi gases, are renormalized by these Landau parameters.

The Fermi surface can be made analogues to an elastic membrane.
The energy cost to deform the Fermi surface can be viewed as
the surface tension, which consists of two contributions from
the kinetic energy and the interaction energy.
The kinetic energy cost is always positive, while the
interaction part can be either positive or negative.
If the Landau parameter $F_l$ is negative and large, {\it i.e.},
$F_l<-(2l+1)$, then the surface tension vanishes in this channel,
and then spontaneous distortion will develop on the Fermi surface
\cite{pomeranchuk1959}.
This class of Fermi surface instability is denoted as Pomeranchuk
instability in the literature.
The simplest Pomeranchuk instability is ferromagnetism
which is an instability in the $s$-wave spin channel.

The Landau interaction function also gives rise to collective excitations
which are absent in free Fermi gases, such as the zero sound mode.
The zero sound is a generalization of the sound waves in fluids
and solids.
In fluids, the sound wave describes the propagation of the density
vibration $\rho(\vec r, t)$, which is a scalar wave; in solids the
sound wave is the vibration of the displacements of atoms from their
equilibrium positions $\vec u(\vec r, t)$, which is a vector wave.
Compared to ordinary fluids which can only support density
fluctuations, Fermi liquid possesses a microscopic structure
of the Fermi surface which can be viewed as an elastic membrane,
whose degree of freedom is infinite described by the spherical
tensor variables $\delta n_{lm}$.
Consider a macroscopically small and microscopically large
volume around $\vec r$, around which a local Fermi surface can
be defined.
The local Fermi surface deformation can vibrate and propagate,
and thus generating sound waves $\delta n_{lm}(\vec r, t)$,
which is the physical picture of the Fermi liquid collective
excitations.
The restoring force for the zero sound arises from Landau interactions
rather than hydrodynamic collisions for the sound modes in ordinary
fluids.

%========================================================
\subsection{Single-particle properties}
\label{sect:fermisurface}

Let us neglect the influence of the confining trap, and also
assume that dipoles polarize along the $z$-axis.
The second quantized Hamiltonian of a single component electric
dipolar fermion system reads
\bea
H_d&=&\sum_{\vec k} [\epsilon(\vec k) -\mu]
c^\dagger(\vec k) c(\vec k)
+ \frac{1}{2V_0}\sum_{\vec k,\vec k^\prime,\vec q}
V_d(\vec q)\nn \\
&\times&
\psi^\dagger(\vec k+\vec q) \psi^\dagger(\vec k^\prime)
\psi(\vec k^\prime+\vec q) \psi(\vec k).
\eea
%The density of states %of two-component Fermi gases
%at the Fermi energy
%is $N_F=\frac{mk_f}{\pi^2\hbar^2}$,
In Sect. \ref{sect:el_dp_FL}, we define the
dimensionless parameter as  $\lambda= d^2 m k_f/(3\pi^2 \hbar^2)$.
It describes the interaction strength, which equals the
ratio between the average interaction energy and the Fermi energy
up to a factor at the order of one.

The Fermi surface structure of an electric dipolar fermion system
is uniform but anisotropic.
Intuitively, the inter-particle distance along the $z$-axis is
shorter than that along $x$ and $y$-axes because the dipolar
interaction is attractive (repulsive) along the $z$-($xy$) direction,
respectively.
Consequently, the Fermi surface will be approximately a
prolate
ellipsoid, elongated along the $z$-axis and compressed in the
equatorial $xy$-plane, which has been investigated
in Refs. [\onlinecite{fregoso2009a,sogo2009,chan2010,miyakawa2008}].

The above picture can be confirmed from the explicit
calculation of the fermion self-energy at the Hartree-Fock level.
The Hartree term vanishes because it involves the spatial
average of the dipolar interaction.
The anisotropy of the Fermi surface can be determined from the
Fock term, while the latter also depends on the actual shape of
the Fermi surface, thus they are coupled together and should be solved
self-consistently.
Nevertheless, at the leading order, we approximate the Fermi surface
as a sphere with the radius in the free space as $k_{f_0}$,
and then $\Sigma^{HF}(\vec k)$ can be calculated analytically
\cite{chan2010} as
\begin{eqnarray}
&&\Sigma^{HF}(\vec{k})=-
2\lambda E_{k_{f_0}} P_2(\cos \theta_k)\
I_{3D} ({k \over k_{f_0}^{3D}}),
\label{eq:HF_3D}
\end{eqnarray}
where $E_{k_{f_0}}=\frac{\hbar^2 k_{f_0}^2}{2m}$, and
$
I_{3D}(x)=  \frac{\pi}{12}\Big \{ 3x^2+ 8
-{3\over x^2}
+\frac{3(1-x^2)^3}{2x^3}
\ln|\frac{1+x}{1-x}| \Big \}$.
In the two-component dipolar Fermi gases, the Hartree term still vanishes
and the Fock term only exists for the intra-component interaction,
thus the Hartree-Fock self-energy remains the same.

%---------------------------------------------------------------
\begin{figure}[tb]
\centering\epsfig{file=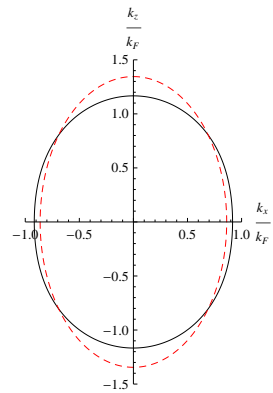,clip=1,
width=0.4\linewidth,angle=0}
\caption{The deformed Fermi surface of the 3D dipolar system
at $\lambda=\frac{1}{2 \pi}$ by the perturbative (solid)
and variational (dashed red) approaches.
The external electric field lies along the $z$-axis.
From Ref. [\onlinecite{chan2010}]. }
\label{fig:fermi_surface_3D}
\end{figure}
%------------------------------------------------------------

The anisotropic Fermi surface distortion is determined by solving the
equation of chemical potential $\mu$ as
\bea
\epsilon_0(\vec
k_f)+\Sigma^{HF}(\vec k_f )=\mu(n,\lambda),
\eea
where $n$ is the particle density.
The Fermi wave vector $\vec k_f$ depends on the polar angle as
\bea
\frac{k_f(\theta_k)}{k_{f_0}}=
1-\frac{4\pi^2}{45}\lambda^2  + \frac{2\pi}{3}\lambda P_2(\cos
\theta_k) \label{eq:fermi_k_3D},
\eea
in which the anisotropic distortion is at the linear order
of $\lambda$, and the $\lambda^2$ term appears to conserve
the particle numbers.
Although Eq. \ref{eq:fermi_k_3D} is only valid at $\lambda\ll 1$,
it provides qualitative features.
The Fermi surface anisotropy was also calculated by using the
numerical variational method in Ref. [\onlinecite{miyakawa2008}].
The comparison between the analytic perturbative result
and the variational one is plotted in
Fig. \ref{fig:fermi_surface_3D} for $\lambda=\frac{1}{2\pi}$.
The Fermi surface based on the first order perturbation result
Eq. \ref{eq:fermi_k_3D} is less prolate than that based on
the variational result.

The anisotropy also exhibits in other single particle quantities.
For example,
the angular distribution of the density of states on the Fermi surface
is calculated as
\bea
N(\Omega_k)\frac{d\Omega_k}{4\pi}&=&\frac{m k_{f0}^{3D}}{\hbar (2\pi)^3}
[1+\frac{5\pi}{3}\lambda P_2(\cos\theta_k)] d\Omega_k, \ \ \,
\label{eq:3DDOS}
\eea
where, $N(\Omega_k)$ is the differential density of states
along the direction of $\Omega_k$.
At the linear order of $\lambda$, $N(\Omega_k)$
develops the same anisotropy of $P_2(\cos\theta)$.
Thus the total DOS at the Fermi energy does not change
compared with that of the free Fermi gas.
Nevertheless, it may be changed due to high order corrections.

%----------------------------------------------------------
\subsection{The Landau interaction matrix}
\label{sect:landau_matrix}

The anisotropic Fermi liquid theory has been constructed in Ref.
[\onlinecite{fregoso2009a,chan2010,rodriguez2014}]
for the dipolar fermion systems.
The anisotropy of the interaction leads to the
mixing among different partial-wave channels, thus we need
to generalize the concept of Landau parameters into the
Landau matrices.

The variation of the Fermi distribution function at momentum $\vec k$
is defined as
\bea
\delta n_{\vec k}=n_{\vec k}-n_{0, k},
\eea
where $n_0(k)=1-\theta(k-k_{f_0}^{3D})$ is the Fermi distribution
function in the absence of interaction.
The ground state energy variation of the single component
dipolar Fermi gas is
\bea
\delta E=\sum_k \epsilon_k \delta n_k+
\frac{1}{2V_0} \lim_{\vec q\rightarrow 0}
\sum_{\vec k, \vec k^\prime} ~   f(\vec k, \vec k^\prime; \vec q)
\delta n_{\vec k, \vec q} \delta n_{\vec k, -\vec q}, \ \ \,
\eea
where, $\vec k, \vec k^\prime$ are momenta close to the Fermi surface;
$f(\vec k, \vec k^\prime; \vec q)$ is the interaction function describing
the forward scattering;
$\vec q$ is the small momentum transfer for the forward scattering
process, which is explicitly kept because of the non-analyticity
of the Fourier component of $V_d(\vec q)$ as $\vec q\rightarrow 0$;
$n_{\vec k, \vec q}=\avg{c^\dagger_{\vec k+\vec q} c_{\vec k}}$,
which is reduced to the fermion occupation number as $\vec q\rightarrow 0$;
$\epsilon_k$ is the renormalized
anisotropic single particle spectra, and at the Hartree-Fock level
$\epsilon_k=\epsilon_k^0+\Sigma_{HF}(\vec k)$.
The Landau interaction function is expressed at the Hartree-Fock level
as
\bea
f(\vec k, \vec k^\prime;\vec q)=V(\vec q)-V(\vec k-\vec k^\prime),
\label{eq:landau_inter}
\eea
where the first and second terms are the Hartree and Fock contributions,
respectively.
Due to the explicit anisotropy, $f(\vec k, \vec k^\prime;\vec q)$ depends
on directions of both $\vec k$ and $\vec k^\prime$, not just the relative
angle between $\vec k$ and $\vec k^\prime$ as in the isotropic Fermi liquids.

The Landau interaction matrix elements for the dipolar system have
been calculated in Ref. [\onlinecite{fregoso2009a}] by Fregoso
{\it et. al.}
According to the Wigner-Eckart theorem, the $d_{r^2-3z^2}$
anisotropy of the dipolar interaction renders the following
spherical harmonics decomposition as
\bea
f(\vec k, \vec k^\prime;\vec q)=\sum_{l,l^\prime;m}
\frac{4\pi f_{ll^\prime;m} }{\sqrt {(2l^+1) (2l^\prime+1)} }
Y_{lm}^*(\Omega_k) Y_{l^\prime m} (\Omega_{\vec k^\prime}),
\nn \\
\label{eq:landauhmncs}
\eea
where $f_{ll^\prime;m}$ remains diagonal for the index $m$ but
couples partial wave channels with $l^\prime=l, l\pm 2$.
The even and odd partial wave channels decouple because of the
even parity of the dipolar interaction.
The $\vec q$ dependence only appear in the channel of
$l=l'=m=0$, in which $f_{00;0}(\vec q)=V_d({\vec q})$.
Other matrix elements at the Hartree-Fock level are tri-diagonal
as
\footnote{The expressions in Eq. \ref{eq:parameter_3D}
we use the
standard normalization convention in Ref. [\onlinecite{leggett1975}]
which is different from that in Ref. [\onlinecite{fregoso2009a}],
thus the parameters in Eq. \ref{eq:fradkin} are modified accordingly.
%Sign errors in the original expressions of Ref. [\onlinecite{fregoso2009}]
%are corrected.
}
\bea
f^{3D}_{ll^\prime;m} &=& d^2 \Big(a_{lm}^{(1)}\delta_{l,l'}
+a_{lm}^{(2)}\delta_{l,l'-2}+a_{l'm}^{(2)}\delta_{l',l-2} \Big), \ \ \,
\label{eq:parameter_3D}
\eea
where
\bea
a_{lm}^{(1)}&=&  \frac{4\pi (l^2+l-3m^2) (2l+1)}{l (l+1)(2l+3)(2l-1)},
\nn \\
a_{lm}^{(2)} &= &-\frac{ 2\pi\sqrt{[(l+1)^2-m^2][(l+2)^2-m^2]} }{(l+1)(l+2)(2l+3)}.
\label{eq:fradkin}
\eea
For each $l\neq 0$, $f^{3D}_{ll^\prime;m}$'s satisfy the sum rule that
\bea
\sum_{m} f_{l l^\prime=l;m}^{3D}=0,
\eea
which reflects the fact the angular average of the dipolar interaction
vanishes.

To make the Landau matrix dimensionless, we multiply the single
component density of states (DOS):
$F^{3D}_{ll^\prime;m}=\frac{\bar m^*}{m}N_0^{3D} f^{3D}_{ll^\prime;m}$,
where $N_0^{3D}= (mk_{f0}^{3D})
/(2\hbar \pi^2)$ is the DOS of free Fermi gas and $m^*$ is the
effective mass.
At the linear order of $\lambda$, $m^*=m$.
For concreteness, some low order Landau matrix elements are
expressed at the linear order of $\lambda$ as
\bea
F_{00;0}(\Omega_{\vec q})&=& 4\pi \lambda P_2(\cos \theta_q), \nn \\
F_{02;0}&=&-\pi  \lambda; \nn \\
F_{22;0}&=&\frac{10\pi}{7}\lambda, \ \ \, \ \ \,
F_{22;\pm1}=\frac{5\pi}{7}\lambda, \nn \\
F_{22;\pm2}&=&-\frac{10\pi}{7}\lambda; \nn \\
F_{11;0}&=&\frac{18\pi}{5} \lambda, \ \ \, \ \ \,
F_{11; \pm1 }=-\frac{9\pi}{5} \lambda;
\nn \\
F_{13;0}&=&-\frac{3\pi}{5}\lambda, \ \ \, ~
F_{13;\pm 1}=-\frac{\sqrt 6 \pi}{5}\lambda; \nn \\
F_{33;0}&=& \frac{14\pi}{15}\lambda, \ \ \, ~~
F_{33;\pm 1}= \frac{7\pi}{10}\lambda; \nn \\
F_{33;\pm 2}&=& 0, \ \ \, \ \ \, \ \ \, ~~~
F_{33;\pm 3}= -\frac{7\pi}{6}\lambda.
\eea

%For the two-component case, the Landau interaction function
%is decomposed into the density channel response $f_s$ and the
%spin-channel response $f_a$ as
%\bea
%f_{\alpha\beta,\gamma\delta;\vec q}(\vec k, \vec k^\prime)
%=f^s(\vec k, \vec k^\prime;\vec q)\delta_{\alpha\beta}\delta_{\gamma\delta}
%+f^a(\vec k, \vec k^\prime;\vec q) \sigma_{\alpha\beta} \sigma_{\gamma\delta},
%\nn \\
%\eea
%where $f_s$ and $f_a$ at the Hartree-Fock level are expressed as
%\bea
%f^s(\vec k,\vec k^\prime;\vec q)&=& V(\vec q)-
%\frac{1}{2} V_d(\vec k-\vec k^\prime); \nn \\
%f^a(\vec k,\vec k^\prime;\vec q)&=& -\frac{1}{2} V_d(\vec k-\vec k^\prime).
%\label{eq:landautwocp}
%\eea
%The DOS of the two-component Fermi gas is multiplied to $f^{s,a}$
%to arrive at the dimensionless Landau matrix elements.
%It is easy to show that they are related to those of the
%single component as
%\bea
%F^{s}_{ll^\prime;m}&=&F^{a}_{ll^\prime;m}=F_{ll^\prime;m},
%\eea
%if at least one of $l$ and $l^\prime$ are nonzero.
%The case of $l=l^\prime=0$ is special, we have
%\bea
%F^s_{00;0}=2 F_{00;0}; \ \ \ F^a_{00;0}=0.
%\eea

%--------------------------------------------------------------------
\subsection{Thermodynamic quantities}
\label{sect:thermo}
The thermodynamic properties, including the anisotropic effective mass and thermodynamic
susceptibilities, are renormalized by the Landau interaction
matrices. %, including the anisotropic effective mass and thermodynamic
%susceptibilities.
For simplicity, only the single-component dipolar systems are
considered here.

%-----------------------------------------------------
\subsubsection{Anisotropic effective mass}
It is well-known that the Galilean %Galiean
invariance leads to the relation
between the bare mass of fermions and the effective mass of
quasiparticles as
\cite{landau1957,landau1959}
\bea
\frac{\vec k}{m}=
\frac{\partial \epsilon(\vec k)}{\partial \vec k}
-\int \frac{d^3 \vec k^\prime}{(2\pi)^3}
f(\vec k; \vec k^\prime) \frac{\partial n(\epsilon (\vec k^\prime) )}
{\partial \vec k^\prime}.
\label{eq:effmass2}
\eea
For an isotropic system, the effective mass $m^*$ is defined as
$\partial \epsilon(\vec k)/\partial \vec k=\vec k/m^*$
for $\vec k$ on the Fermi surface.
The renormalization of $m^*$, or, the renormalization of the
density of states, is
$\frac{m^*}{m}=1+\frac{1}{3}F^s_1$,
which affects the specific heat as $C_{FL}/C_{FG}=m^*/m$,
with $C_{FL}$ and $C_{FG}$ specific heats for the Fermi liquid and
ideal Fermi gas, respectively.

The dipolar Fermi gas is Galilean invariant so that Eq. \ref{eq:effmass2}
is still valid.
However, due to the anisotropy, a self-consistent solution has
to be done numerically.
To the linear order of $\lambda$, we approximate
$\epsilon(\vec k^\prime)$ in the right-hand-side of Eq. \ref{eq:effmass2}
with the free fermion energy.
Defining the radial effective mass as
$m^*_{3D,\pp} (\theta_k)=\frac{1}{k_f(\theta_k)}
[\hat k \cdot \frac{\partial \epsilon(\vec k)}{\partial \vec k}]$,
we arrive at
\bea
\frac{1}{m}
&=& \frac{1}{ m^*_{3D,\pp} (\theta_k)}
+\frac{1}{m}\Big[\tilde F_{11,\pp}(\theta_k) + \tilde F_{13,\pp}(\theta_k) \Big ],
\label{eq:effmass3}
\eea
where $\tilde F_{11,\pp}(\theta_k)$ and $\tilde F_{13,\pp}(\theta_k)$
are the angular dependent Landau parameters defined as follows
\bea
\label{eq:f11q}
\tilde F_{11,\pp}(\theta_k)&=& \frac{4\pi}{3}\sum_m \frac{F_{11;m}}{3}
|Y_{lm} (\theta_k,0)|^2
%=\frac{18\pi \lambda}{5} P_2(\cos \theta_k),
\nn \\
\tilde F_{13,\pp}(\theta_k)&=&
\frac{4\pi}{3}\sum_{m=0,\pm1} \frac {F_{13;m}}{\sqrt {21}}
Y^*_{3m}(\theta_k,0) Y_{1m}(\theta_k,0). \nn
%&=& \frac{\sqrt{21}\pi}{5} \lambda P_2(\cos \theta_k).
\eea
Thus to the linear order of $\lambda$, the anisotropic
radial effective mass is
\bea
\frac{1}{m_{3D,\pp}^*(\theta_k)}
=\frac{1-\pi\lambda P_2(\cos\theta_k)}{m}.
\eea

%-----------------------------------------------------
\subsubsection{Thermodynamic susceptibilities}

Viewing the Fermi surface as an elastic membrane, we define the angular
variation of the fermion distribution as
\bea
\delta n (\Omega_{\vec k})=\int\frac{k^2 d k}{(2\pi)^3} \delta n_{\vec k},
\eea
which can be further expanded in terms of the spherical
harmonics as
\bea
\delta n(\Omega_{\vec k})=\sum_{lm} Y_{lm} (\Omega_{\vec k})
\delta n_{lm}.
\label{eq:nlm}
\eea

For a Fermi surface distortion characterized by a set of
$\delta n_{lm}$, the energy variation is calculated as
\bea
\frac{\delta E}{V}
&=& 4\pi \Big\{ \frac{1}{2\chi_0}\sum_{ll^\prime;m}\delta n_{lm}^* \delta
n_{l^\prime m} K_{ll^\prime;m}^{3D} -h_{lm} \delta n_{lm} \Big\}, \ \ \, \ \ \,
\label{eq:FLenergy}
\eea
where $\chi_0=\frac{\bar m^*}{m}N_0$ is the density of states at the
Fermi energy; $\bar m^*$ is the average value of the effective mass
on the Fermi surface which equals $m$ at the linear order of $\lambda$;
$h_{lm}=h_{lm}^{ex}+h_{20}^0$.
$h_{lm}^{ex}$ is the external field in the partial wave channel of $lm$,
and $h_{20}^0=\frac{2}{3}\sqrt{\frac{\pi}{5}}\lambda E_{k_{f0}}^{3D}$
is the explicit symmetry breaking arising from the dipolar interaction.
The matrix kernel $K_{ll^\prime}$ in Eq. \ref{eq:FLenergy} contains
two parts as
\bea
K_{ll^\prime;m}= M_{ll^\prime;m}
+\frac{F_{ll^\prime;m}^{3D}}{\sqrt{(2l+1)(2l^\prime+1)}},
\label{eq:m_matrix_3D}
\eea
in which, $M_{ll^\prime;m}$ is the kinetic energy contribution.
At the linear order of $\lambda$, $M_{ll^\prime;m}$ is calculated
as
\begin{eqnarray*}
M_{ll^\prime;m}^{3D}= m_{lm}^{(1)} \delta_{ll^\prime}+  m_{lm}^{(2)} \delta_{l, l^\prime-2}
+ m_{lm}^{(2)} \delta_{l,l^\prime-2},
\end{eqnarray*}
where
\bea
m_{lm}^{(1)} &=&1+\frac{l(l+1)}{4(2l+1)}a^{(1)}_{lm}\lambda , \nn \\
m_{lm}^{(2)} &=& \frac{3(l+1)(l+2)}{4\sqrt{(2l+1)(2l+5)}}a^{(2)}_{lm}\lambda.
\eea

The expectation value of $\delta n_{lm}$ in the field of $h_{lm}$
can be straightforwardly calculated as
\bea
\delta n_{lm} =\chi_0  (K^{3D}_{ll^\prime;m})^{-1} h_{l^\prime m}.
\label{eq:renormalized_suscept}
\eea
Thus $\chi_0 (K^{3D}_{ll^\prime;m})^{-1}$ is the renormalized susceptibility
matrix for the 3D dipolar Fermi system \cite{chan2010}.

%----------------------------------------------------------------
\subsubsection{Pomeranchuk stabilities}
The inverse of an eigenvalue of the matrix $K_{ll^\prime;m}$
can be considered as a thermodynamical susceptibility in the
corresponding eigen-channel.
If all the eigenvalues of  $K_{ll^\prime;m}$ are positive, {\it i.e.},
this matrix is positive-definite, then the system is thermodynamically
stable.
If any of them becomes negative, the corresponding susceptibility
diverges, which signals the onset of the Fermi surface instability
of the Pomeranchuk type \cite{fregoso2009,chan2010}.
For the isotropic system, $K_{ll^\prime;m}$ is already diagonal and
$m$-independent, and the criterion for the Pomeranchuk instability
is the well-known one
\bea
F_l<-(2l+1).
\eea

For the anisotropic dipolar system, the $K_{ll^\prime;m}$ matrix
needs to be diagonalized to analyze its thermodynamic instabilities.
The two strongest instabilities lie in the sectors of of $m=0$ and
$m=\pm 2$, and $l$'s are even.
For the case of $m=0$, $F_{00;0}(\vec q)$ is singular as
$\vec q\rightarrow 0$, explicitly depending on the direction
of $\vec q$.
The most negative eigenvalue occurs when $\vec q$ lies
in the equatorial plane.
The corresponding eigenvector mainly lies in the $s$-wave channel
with $l=0$, and the eigenvalue $\mu_s$ reaches zero at $\lambda_s=0.135$.
This instability corresponds to the Fermi surface collapsing
with a density modulation wave vector in the equatorial plane.
This result nicely agrees with %by
the numerical calculation in Ref.
[\onlinecite{ronen2010}], in which the onset of an unstable collective
mode implies the Fermi surface collapsing starting from
$\lambda_s\sim0.14$.
It should be noted that, actually, this instability is mostly driven by
the Hartree term interaction $V_d(\vec q)$, which cannot
be simply dropped off by setting $\vec q=0$. %, o
Otherwise, the stability of the dipolar Fermi gas
would be significantly overestimated.

As for the sector of $m=\pm 2$, the eigenvectors of the minimal eigenvalues
mainly lie in the $d_{x^2-y^2\pm 2 ixy}$-wave channels.
The lowest eigenvalues $\mu_{d_{\pm 2}}$ touch zero at
$\lambda_{d_{\pm2}}=0.35$.
This instability corresponds to the biaxial nematic
instability of the Fermi surfaces studied in Ref. [\onlinecite{fregoso2009}].
With the purely dipolar interaction, the $s$-wave channel instability
occurs before the $d$-wave channel one because
$\lambda_{d_\pm2}>\lambda_{s}$.
Nevertheless, the $s$-wave channel instability can be cured by introducing
a positive non-dipolar short-range $s$-wave scattering potential $V_{00;0}$,
which adds to the Landau parameter of $F_{00;0}^{3D}$ without affecting
other channels.

%--------------------------------------------------------
\subsection{The collective zero sound mode}
\label{sect:zero_sound}

In this part, we review the calculation of the zero-sound-like
collective mode in the dipolar Fermi gases
\cite{ronen2010,chan2010}.
The anisotropic dipolar interaction brings a new feature: the zero
sound excitation can only propagate within a certain range of
directions beyond which the sound mode is damped.

%--------------------------------------------------------------
\subsubsection{Generalized dynamical response functions}
\label{subsect:generalized_response}
The Boltzmann equation for the collective excitation of the
single-component Fermi liquid is \cite{negele1988}
\bea
\frac{\partial}{\partial t} n(\vec r, \vec k, t)
&+&\sum_i  \frac{\partial \epsilon (\vec r, \vec k, t)}{\partial k_i}
\frac{\partial n (\vec r, \vec k, t)}{\partial r_i} \nn \\
&-&
\sum_i \frac{\partial \epsilon (\vec r, \vec k, t)}{\partial r_i}
\frac{\partial n (\vec r, \vec k, t)}{\partial k_i}
=0,
\eea
where $n(\vec r, \vec k, t)$ and  $\epsilon(\vec r, \vec k, t)$
are the density and energy distributions in the phase space.

In order to linearize the Boltzmann equation, the small variations
of $n(\vec r, \vec k, t)$ and $\epsilon(\vec r, \vec k, t)$ are defined
as
\bea
n(\vec r, \vec k, t)&=&n_{0,d}(k)+\delta \nu(\vec r, \vec k, t),\nn \\
\epsilon(\vec r, \vec k, t)&=&\epsilon_{HF}(k)
+\int \frac{d^3 k^\prime}{(2\pi)^3}
f(\hat k, \hat k^\prime) \delta \nu (\vec r,\hat k^\prime, t), \ \ \,
\eea
where $\delta \nu$ is defined with respect to the deformed
equilibrium Fermi surface,
and $\epsilon_{HF}$ is the Hartree-Fock single particle spectrum.
Substituting $\delta \nu(\vec r, \vec k, t)
=\sum_q \delta \nu_{\vec k} e^{i (\vec q \cdot \vec r-\omega t)}$,
the linearized Boltzmann equation is arrived at
\bea
\delta \nu(\Omega_k)-   \frac{ N(\Omega_k) \vec v_{\vec k} \cdot \vec k}
{\omega - \vec v_{\vec k} \cdot \vec q}
\int \frac{d\Omega_{k^\prime}}{4\pi} f_{\vec k \vec k^\prime}
\delta \vec \nu(\Omega_{k^\prime}) =0, \ \ \,
\eea
where $\vec v_{\vec k}$ is the Fermi velocity;
$N(\Omega_k)$ is the differential density of states
defined in Eq. \ref{eq:3DDOS};
$\delta \nu (\Omega_k)$ is defined as
\bea
\delta \nu (\Omega_k)=\int \frac{k^2 dk}{(2\pi)^3} \delta \nu(\vec k).
\eea
The spherical harmonics decomposition can be performed as
\bea
\sum_{l'm'} \Big\{\delta_{ll'}\delta_{mm'}+ \sum_{l^{\prime\prime}}
\chi_{ll^{\prime\prime};mm'}(\omega, \vec q)
F_{l^{\prime\prime}l';m'}   \Big\}\delta  \nu_{l'm'}=0, \ \ \
\label{eq:boltzmann_3D}
\eea
where $\delta \nu_{lm}$ is the component in terms of
the spherical harmonics, and,
\bea
\chi_{ll';mm'}(\omega,\vec q) &=& -\frac{1}{\sqrt {(2l+1)(2l'+1)}}
\int d \Omega_p \frac{N^{3D}(\Omega_{\vec p})}{N^{3D}_0} \nn \\
&\times& Y^*_{lm}(\Omega_{\vec p})
\frac{\vec v_p^{3D} . \vec q}{\omega - \vec v_p^{3D} . \vec q}
Y_{l'm'}(\Omega_{\vec p}).
\label{eq:response_3D}
\eea
Due to the dipolar anisotropy, for a general propagation
direction of $\vec q$, $\delta \nu_{lm}$ in different channels are coupled.

%--------------------------------------------------------
\subsubsection{The $s$-wave channel approximation}
We first truncate Eq. \ref{eq:response_3D} by only keeping the
$s$-wave channel of $l=0$.
Even at this level, the anisotropy of the zero sound mode
has already appeared.
Taking into account the Hartree-Fock single particle spectra and
the anisotropic Fermi surface, $\chi_{00;00}(\omega, \vec q)$
is given by
\bea
\chi_{00;00}(\omega,\vec q)&=& 1 -\int \frac{d\Omega_k}{4\pi}
\frac{N(\Omega_k)}{N_0}
\frac{s}{s-f(\Omega_k,\Omega_q)+i\eta}, \nn \\
\eea
where $f(\Omega_k, \Omega_q)=
\frac{\vec q \cdot \nabla_k \epsilon_{HF}(\vec k)}{v_{f_0} q}$,
and the propagation direction $\vec q$ is chosen in the $xz$-plane
with the polar angle $\theta_q$.
Then the zero sound mode dispersion is determined by
\bea
1+F_{00;0}(\Omega_{\vec q})\chi_{00;00}(\omega,\vec q)=0.
\eea

\begin{figure}[tb]
\centering\epsfig{file=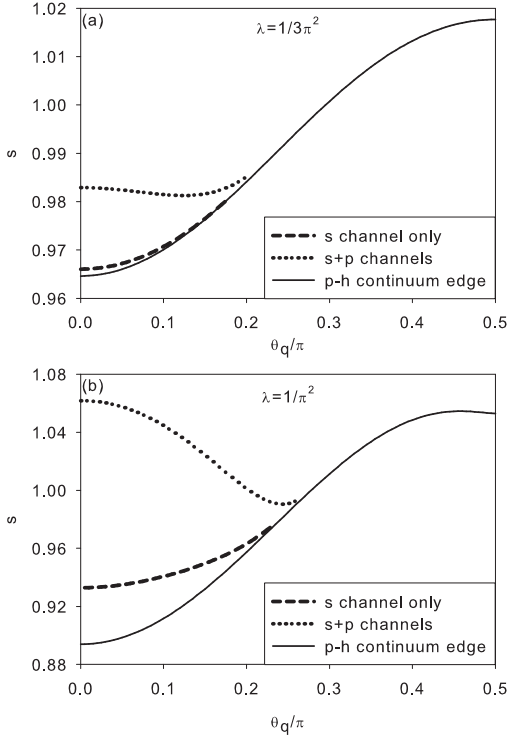,width=0.8\linewidth,angle=0}
\caption{Dispersions of the zero sound $s(\theta_q)=\omega
(\theta_q)/v_{f0}^{3D} q$ for the pure dipolar interaction
at $a)$ $\lambda=1/(3\pi^2)$ and $b)$ $\lambda=1/\pi^2$.
When the sound speed hits the particle hole continuum,
the sound becomes damped.
These results are in good agreement with the numerical study
in Ref. [\onlinecite{ronen2010}].
%(The parameters are chosen for a better comparison with
% the Ref. [\onlinecite{ronen2010}], where their parameters
%$D=3\pi^2\lambda$ are chosen to be 1 and 3.)
(From Ref. [\onlinecite{chan2010}].) }
\label{fig:zero_3D_pwave}
\end{figure}

The quantity $s(\theta_q)=\omega(\theta_q)/(v_{f_0} q)$
is defined to represent the angular dependent zero sound dispersion, which
is solved numerically and plotted in Fig. \ref{fig:zero_3D_pwave}
along with the edge of particle-hole continuum.
The zero sound propagation angle is restricted and its dispersion
$s(\theta_q)$ is anisotropic.
For large angles of $\theta_q$, the sound excitation enters the
particle-hole continuum and is thus damped.

%----------------------------------------------------------------
%---------------------------------------------------------------

%-------------------------------------------------------
\subsubsection{Correction from the coupling to the $p$-wave
longitudinal channel}

Even in the isotropic Fermi liquid state, because the propagation
direction $\vec q$ of the zero sound already breaks the 3D rotational
symmetry to the uni-axial one,  actually the zero sound mode mixes
all the longitudinal channels of $\delta n_{l0}$.
If the Landau parameter $F_1$ is not small compared to $F_0$,
the mixing between the $s$ and $p$-wave longitudinal channels
significantly modifies the sound velocity.
In the isotropic Fermi liquid state, the modified sound velocity
is determined by the following equation as  \cite{negele1988}
\bea
\frac{-1 }{ F_{0}+
\frac{s^2 F_{1} }{1+\frac{F_{1} }{3} } }
= 1-\frac{s}{2} \ln |\frac{1+s}{1-s}| .
\label{eq:zerospectra}
\eea
For example, in the $^3$He system at 0.28 atm, if only considering the
$s$-wave channel, the sound velocity is calculated as
$s=v_s/v_f=2.0$ based on $F_0^s=10.8$.
After including the coupling of the $p$-wave longitudinal
channel in which $F_1^s=6.3$, the revised value of $s$
increases to 3.6 in agreement with the experimental
measurements \cite{negele1988}.

The case of the dipolar Fermi gas is more complicated.
If the propagating direction $\vec q$ is not along the $z$-axis,
no rotational symmetry is left, and thus, in principle,
the longitudinal and transverse $p$-wave components are mixed.
Here, the spherical harmonic functions ${\tilde Y}_{l=1,m=0}$
(longitudinal) and ${\tilde Y}_{l=1,m=\pm 1}$ (transverse)
are defined according to the principle axis along $\vec q$
instead of the $z$-axis.
Nevertheless, usually the transverse $p$-wave channel mode
is overdamped unless the $p$-wave channel Landau parameter
is positive and large, thus their effect to the zero sound
mode is small and will be neglected.

By keeping the mixing between the $s$-wave and the longitudinal
$p$-wave modes, Eq. \ref{eq:response_3D} is reduced to a
$2\times 2$ matrix equation, and the collective mode
can be solved based on
\bea
\mbox{Det} [ 1 + N(\vec q, \omega) ]=0,
\eea
where the matrix kernel of $N (\vec q,\omega)$ reads
\bea
N(\vec q, \omega)=
\left( \begin{array}{cc}
\chi_{00;00} (\vec q, \omega) F_{00;0} (\vec q)
& \tilde\chi_{10;00} (\vec q, \omega)
\frac{\tilde F_{110} (\vec q)}{3} \\
\tilde \chi_{10;00} (\vec q,\omega) F_{00;0}  (\vec q)
& \tilde \chi_{11;00}(\vec q, \omega)
\frac{\tilde F_{11;0} (\vec q)}{3}
\end{array}
\right ). \nn \\
\eea
$\tilde F_{11;0}^{3D} (\vec q)$ is the longitudinal $p$-wave Landau
parameter defined as
\bea
\tilde F_{11;0} (\vec q)
&=&\cos^2\theta_q F_{11;m=0}+\sin^2\theta_q F_{11;m=\mp 1},
%&=&\frac{18\pi}{5}\lambda P_2(\cos\theta_q),
\eea
and the response functions are
\bea
\tilde \chi_{10;00}(\vec q, \omega)
&=&- \sqrt{3} \int \frac{d\Omega_k}{4\pi}
\frac{ N(\Omega_k)}{N_0}\frac{  (\hat q \cdot \hat k)
f(\Omega_k, \Omega_q)}
{s-f(\Omega_k, \Omega_q)}, \nn \\
\tilde \chi_{11;00}(\vec q, \omega)&=&- 3
\int \frac{d\Omega_k}{4\pi} \frac{ N(\Omega_k)}{N_0}
\frac{  (\hat q \cdot \hat k)^2 f(\Omega_k, \Omega_q)}
{s-f(\Omega_k, \Omega_q)}, \nn
\eea
where $\vec q$ lies in the $xz$-plane with the polar angle $\theta_q$,
and $\hat q \cdot \hat k=\sin\theta_q \sin\theta_k \cos\phi_k
+\cos\theta_q \cos\theta_k$.

The numeric solution taking into account the anisotropic Fermi
velocity and Fermi surface is performed, and the zero sound velocity
as a function of $\theta_q$ is plotted in Fig. \ref{fig:zero_3D_pwave}.
The longitudinal $p$-wave mode modifies the sound velocity dispersion
significantly.
These results are in a good agreement with a fully numerical calculation
based on the same Boltzmann transport theory \cite{ronen2010}.
This indicates that the zero sound mode is well captured by
the coupling between the $s$-wave and longitudinal $p$-wave channels.

%%%%%%%%%%%%%%%%%%%%%%%%%%%%%%%%%%%%%%%%%%%%%%%%%%%%%%%%%%%%%%
\section{The SO coupled Fermi liquid theory of the magnetic
dipolar fermions}
\label{sect:mg_dp_FL}

The magnetic dipolar interaction brings a new ingredient to the
Fermi liquid theory, i.e.,  the SO coupled nature
\cite{fujita1987,fregoso2009,fregoso2010,li2012,sogo2012,li2012a,
bhongale2013,tohyama2013,ashrafi2013}.
Certainly, for the experimental system of the $^{161}$Dy atoms whose
hyperfine spin is as large as  $F=\frac{21}{2}$ \cite{lu2010,lu2010a,lu2012},
%and $^{167}$Er with 7 $\mu_B$ \cite{aikawa2014,aikawa2014a,aikawa2014b},
the theoretical analysis on magnetic dipolar interactions
will be very challenging.
Nevertheless, the spin-$\frac{1}{2}$ case exhibits nearly all
the qualitative features of the magnetic dipolar interaction,
and thus will be used as a prototype model below.

If the magnetic dipolar systems are partially polarized, this
SO coupling already appears at the single-particle level exhibiting
anisotropic Fermi surfaces, which is a result from the ferro-nematic
coupling as shown by Fregoso {\it et. al.} \cite{fregoso2009,fregoso2010}.
For the unpolarized case, the Fermi surfaces remain spherical
without splitting.
Nevertheless, the effects of the SO coupling appear at the
interaction level \cite{li2012a,sogo2012}, including Fermi surface
Pomeranchuk instabilities, and SO coupled collective modes.

The second quantized  Hamiltonian of the spin-$\frac{1}{2}$ fermions
with the magnetic dipolar interaction is expressed as
\bea
H_{md}&=&\sum_{\vec k,\alpha} [\epsilon(\vec k) -\mu]
c^\dagger_\alpha(\vec k) c_\alpha(\vec k)
+\frac{1}{2V_0}\sum_{\vec k,\vec k^\prime,\vec q}
V_{\alpha\beta;\beta^\prime\alpha^\prime}(\vec q)
\nn \\
&\times& \psi^\dagger_\alpha(\vec k+\vec q) \psi^\dagger_\beta(\vec k^\prime)
\psi_{\beta^\prime}(\vec k^\prime+\vec q) \psi_{\alpha^\prime}(\vec k).
\eea
Similarly, the dimensionless interaction parameter can be
defined accordingly as $\lambda_{m}= \mu_B^2g_F^2 mk_f/(\pi^2\hbar^2)$.

The SO coupled Landau interaction function is reviewed
and the SO coupled partial wave decomposition is performed
in Sect. \ref{sect:landau};
the Pomeranchuk instability is reviewed
in Sect. \ref{sect:soPomeranchuk};
the zero sound-like excitation with the SO coupled feature
is reviewed in Sect. \ref{sect:socoll}.

%-----------------------------------------------------------------
\subsection{The SO coupled Landau interaction}
\label{sect:landau}

In Ref. [\onlinecite{sogo2012}], the Landau interaction function of
the magnetic dipolar fermions with a general hyperfine-spin $F$ was given.
For simplicity and concreteness, below we still use the spin-$\frac{1}{2}$
case for illustration \cite{li2012a}.
Based on the Fourier transform of the magnetic dipolar interaction
Eq. \ref{eq:four_mgdp}, the Landau function of the spin-$\frac{1}{2}$
magnetic dipolar system is expressed at the Hartree-Fock level as
\bea
f_{\alpha\alpha^\prime,\beta\beta^\prime}(\vec k, \vec k^\prime;\vec q)
&=&f^{H}_{\alpha\alpha^\prime,\beta\beta^\prime}(\hat q)
+f^{F}_{\alpha\alpha^\prime,\beta\beta^\prime}(\vec k,\vec k^\prime)\nn \\
&=& \frac{\pi g_F^2\mu_B^{2}}{3}
\big(M_{\alpha\alpha^\prime,\beta\beta^\prime} (\hat q)
- M_{\alpha\alpha^\prime,\beta\beta^\prime} (\hat p)\big), \nn
\label{eq:landau_mag}
\eea
where $\vec k$ and $\vec k^\prime$ are at the Fermi surface;
the small momentum transfer $\vec q$ in the Hartree term
is explicitly kept due to the singularity at $\vec q\rightarrow 0$
in Eq. \ref{eq:four_mgdp};
$\hat p$ is the unit vector defined as
$\hat p= \frac{\vec k -\vec k^\prime}{|\vec k -\vec k^\prime|}$;
the matrix kernel $M_{\alpha\alpha^\prime,\beta\beta^\prime}(\hat m)$
only depends on the direction of $\hat m$ as
\bea
M_{\alpha\alpha^\prime,\beta\beta^\prime} (\hat m)
=3(\vec \sigma_{\alpha\alpha^\prime} \cdot \hat m)(\vec
\sigma_{\beta\beta^\prime} \cdot \hat m)
-\vec \sigma_{\alpha\alpha^\prime} \cdot \vec \sigma_{\beta\beta^\prime},
\ \ \,
\eea
for $\hat m=\hat p$ and $\hat q$.
In order to arrive at $f^F_{\alpha\alpha^\prime,\beta\beta^\prime}(\vec k,
\vec k^\prime)$ in Eq. \ref{eq:landau_mag}, the following
identity is used
\bea
&&3(\vec \sigma_{\alpha\beta^\prime} \cdot \hat p)
(\vec \sigma_{\beta\alpha^\prime} \cdot \hat p)
-\vec \sigma_{\alpha\beta^\prime}\cdot \vec \sigma_{\beta\alpha^\prime}
\nn \\
&=&3(\vec \sigma_{\alpha\alpha^\prime} \cdot \hat p)
(\vec \sigma_{\beta\beta^\prime} \cdot \hat p)
-\vec \sigma_{\alpha\alpha^\prime} \cdot \vec \sigma_{\beta\beta^\prime}.
\eea

%-------------------------------------------------------
\subsubsection{The SO partial-wave decomposition}

It is convenient to work in the SO coupled bases for the magnetic
dipolar Fermi liquid theory.
The variation of the single particle density matrix in momentum
space is defined as
$\delta n_{\alpha\alpha^\prime}(\vec k)= n_{\alpha\alpha^\prime}(\vec k)
-\delta_{\alpha\alpha^\prime}n_0(\vec k)$, where
$n_{\alpha\alpha^\prime}(\vec k)=\avg{\psi^\dagger_{\alpha} (\vec k)
\psi_{\alpha^\prime}(\vec k)}$ and $n_0(\vec k)$ refers to the
ground state distribution of the free Fermi system.
As for spin indices,
$\delta n_{\alpha\alpha^\prime}(\vec k)$ can be expanded as
\bea
\delta n_{\alpha\alpha^\prime}(\vec k)=\sum_{Ss_z} \delta n_{Ss_z}(\vec k)
\chi_{Ss_z,\alpha\alpha^\prime},
\eea
where $\chi_{Ss_z,\alpha\alpha^\prime}$ are the bases for the
particle-hole singlet (density) channel with $S=0$
and triplet (spin) channel with $S=1$, respectively,
defined as
\bea
\chi_{10,\alpha\alpha^\prime}&=& \sigma_{z,\alpha\alpha^\prime}, ~~~
\chi_{1\pm 1,\alpha\alpha^\prime}=\frac{\mp1}{\sqrt 2}
(\sigma_{x,\alpha\alpha^\prime} \pm i\sigma_{y,\alpha\alpha^\prime}),\nn \\
\chi_{00,\alpha\alpha^\prime}&=&\delta_{\alpha\alpha^\prime}.
\eea

Similarly as before, we integrate $\delta n_{\alpha\alpha^\prime}(\vec k)$
along the radial direction, and arrive the angular distribution
$\delta n_{\alpha\alpha^\prime}(\hat k)$.
In the SO decoupled bases, it is expanded as
\bea
\delta n_{\alpha\alpha^\prime}(\hat k)
=\sum_{LmSs_z} \delta n_{LmSs_z} Y_{Lm}(\hat k)
\chi_{Ss_z,\alpha\alpha^\prime}.
\eea
More conveniently, it can be reorganized in the SO coupled bases as
\bea
\delta n_{\alpha\alpha^\prime}(\hat k)
&=&\sum_{JJ_z;LS} \delta n_{JJ_z;LS} ~
{\cal Y}_{JJ_z;LS}(\hat k, \alpha\alpha^\prime),
\eea
where
$\delta n_{JJz;LS}=\sum_{ms_z} \avg{LmSs_z|JJ_z}
\delta n_{LmSs_z}$; ${\cal Y}_{JJz;LS}(\hat k, \alpha\alpha^\prime)$
is the SO coupled spherical harmonic functions
\bea
{\cal Y}_{JJz;LS}(\hat k, \alpha\alpha^\prime)
&=&\sum_{ms_z}\avg{LmSs_z|JJ_z} Y_{Lm}(\hat k)
\chi_{Ss_z,\alpha\alpha^\prime}. \nn
\label{eq:socouple}
\eea

Using the SO coupled bases,
the Landau matrix is diagonal with
respect to the total angular momentum $J$ and its $z$-component
$J_z$ as
\bea
\frac{N_0}{4\pi } f_{\alpha\alpha^\prime;\beta\beta^\prime}(\hat k, \hat k^\prime)&=&
\sum_{JJ_z L L^\prime}
{\cal Y}_{JJz;L1}(\hat k,\alpha\alpha^\prime)
 F_{JJ_z L1;JJ_z L^\prime 1} \nn \\
&\times&
{\cal Y}^\dagger_{JJz;L^\prime 1}(\hat k,\beta\beta^\prime).
\eea
The matrix kernel $F_{JJ_zL1;JJ_zL^\prime1}$ reads
\begin{widetext}
\bea
F_{JJ_zL1;JJ_zL^\prime 1}&=&\frac{\pi\lambda}{3}\delta_{J,1}\delta_{L,0}
\delta_{L^\prime,0}
(2 \delta_{J_z,0}-\delta_{J_z, \pm1}) +
\sum_{ms_z; m^\prime s_z^\prime}
\avg{Lm1s_z|JJ_z}
\avg{L^\prime m^\prime 1s_z^\prime|JJ_z}
T^{F}_{Lm1s_z;L^\prime m^\prime 1s^\prime_z},
\label{eq:LL_para}
\eea
in which the first term is the Hartree contribution with $\hat q$
set as the $z$-axis; the second term is the Fock contribution with
$T^{F}_{Lm1s_z;L^\prime m^\prime 1s^\prime_z}$ defined in the SO
decoupled bases as \cite{sogo2012,li2012a}
\bea
T^{F}_{Lm1s_z;L^\prime m^\prime 1s^\prime_z}&=&
-\frac{\pi \lambda}{2}
\Big(\frac{\delta_{LL^\prime}}{L(L+1)}
-\frac{\delta_{L+2,L^\prime}}{3(L+1)(L+2)}
-\frac{\delta_{L-2,L^\prime}}{3(L-1)L}\Big)
\int d \Omega_r \Big( \delta_{s_zs_z^\prime}
-4\pi Y_{1s_z}(\Omega_r) Y^*_{1s_z^\prime}(\Omega_r)\Big)\nn \\
&\times&
Y_{Lm}(\Omega_r) Y_{L^\prime m^\prime}^* (\Omega_r).
\label{eq:mgdr_landau}
\eea
\end{widetext}

The same value of $J$ may arise from $L$ with different parities.
Below we use $(J^{\pm} J_zLS)$ to represent different angular
momentum channels, where $\pm$ is the parity eigenvalue.
The Hartree term of Eq. \ref{eq:LL_para} only contributes
to  the $(1^+J_z01)$ sector, which explicitly depends on
$J_z$ because $\vec q$ (chosen as the $\hat z$-axis)
breaks the 3D rotation symmetry down to the uniaxial rotation symmetry.
For other sectors, $F_{J^\pm J_zL1;J^\pm J_zL^\prime1}$ does not depend
on $J_z$ as required by the Wigner-Eckart theorem.
The matrix $F_{JJ_zL1;JJ_zL^\prime1}$ is nearly diagonalized except
for the case with $L\neq L^\prime$.
For $J^+=0^+,2^+,..$ and $J^-=1^-,3^-,...$,
there is only one possibility that $L=J$,  and thus
$F_{J^\pm J_zL1;J^\pm J_zL^\prime1}$ is already diagonalized.
In comparison, for $J^+=1^+,3^+,...$ and $J^-=0^-, 2^-,...$
$L$ can take two different values of $L=J\pm 1$, and thus
the matrix is reduced to $2\times 2$ (For $J^-=0^-$
the only possibility is $L=1$.).
The concrete forms of  $F_{J^\pm J_zL1;J^\pm J_zL^\prime1}$
for a few low orders of $J^\pm$ are given in
Ref. [\onlinecite{li2012a,sogo2012}].

%-----------------------------------------------------------
\subsubsection{Thermodynamics susceptibilities}

The variation of the ground state energy in the
SO coupled bases is expressed  as
\bea
\frac{\delta E}{V}
&=&16\pi\Big\{ \frac{1}{2\chi_0} \sum_{JJ_z L L^\prime S}
\delta n^*_{JJ_z;LS} M_{JJ_zlS; JJ_z L^\prime S} \delta n_{JJ_z;L^\prime S} \nn \\
&-& \sum_{JJ_z L S} h_{JJ_zLS} \delta n_{JJ_z;L S}
\Big\},
\eea
where the matrix kernel is
\bea
M_{JJ_zLS; JJ_z L^\prime S} =\delta_{LL^\prime} + F_{JJ_zLS;JJ_z L^\prime S};
\label{eq:matrix}
\eea
$\chi_0=N_0$ is the Fermi liquid density of states;
$h_{JJ_zLS}$ is the external field.
At the Hartree-Fock level, $N_0$ receives no renormalization from the
magnetic dipolar interaction.
The expectation value of $\delta n_{JJ_zLS}$ is calculated as
\bea
\delta n_{JJ_zLS}=\chi_0 \sum_{L^\prime}(M)^{-1}_{JJ_zLS;JJ_z L^\prime S}
h_{JJ_zL^\prime S}.
\label{eq:nh}
\eea
The thermodynamic stability condition is equivalent to that all
the eigenvalues of the matrix $M_{JJ_zLS;JJ_zL^\prime S}$ are positive.

Fregoso {\it et al.} \cite{fregoso2009,fregoso2010}
found that a uniform magnetic
field  $\vec h$ along the $z$-axis not only induces spin polarization,
but also a spin-nematic order in the channel of $(J^+J_zLS)=(1^+021)$.
As a result, the external magnetic field induces an effective
SO coupling
\bea
H_{hso}&=& \frac{\sqrt 2 }{12} \pi \lambda h  \sum_{k}
\psi^\dagger_{\alpha}(\vec k) \Big\{ %\frac{1}{\sqrt{8 \pi}}
\big[(k^2-3k_z^2) \sigma_z  \nn \\
&-& 3 k_z (k_x \sigma_x + k_y \sigma_y) \big] \Big\}
\psi_\beta (\vec k).
\label{eq:so_1}
\eea
Apparently, Eq. \ref{eq:so_1} breaks TR symmetry, which is
markedly different from the relativistic SO coupling
in solids.

%-----------------------------------------------------
\subsection{SO coupled Pomeranchuk instabilities}
\label{sect:soPomeranchuk}
If one of the eigenvalue of the Landau matrix becomes negative,
Pomeranchuk instability occurs in the corresponding channel.
The SO coupled nature in the magnetic dipolar system
manifests as follows.

Let us first consider the channel of $J=1^+$ \cite{li2012a}.
In the absence of external fields, a density wave with a long wave
length $q\rightarrow 0$ can take the advantage of the Hartree channel
interaction.
The leading instability lies in the sector with $J_z=\pm 1$.
Without loss of generality, we use the $J_z=1$ sector as an example:
\bea
\left(
\begin{array}{cc}
F_{1101;1101}& F_{1101;1121}\\
F_{1121;1101}& F_{1121;1121}
\end{array}
\right)
=
\frac{\pi \lambda_m}{12}
\left(
\begin{array}{cc}
-4 & \sqrt{2} \\
 \sqrt{2} & 1
\end{array}
\right),
\label{eq:landau_2}
\eea
whose negative eigenvalue and the associated eigenvector are
\bea
w^{1^+1}_1 &=& -0.37 \pi\lambda_m, \ \ \, \psi^{1^+ 1}_1=(0.97,-0.25)^T.
\label{eq:eigen_2}
\eea
The instability occurs at $w_1^{1^+1}<-1$, or, equivalently,
$\lambda_m>\lambda_{1^+1}^c=0.86$, and
the eigenvector shows that it is nearly a ferromagnetic instability
hybridized with a small component of the ferro-nematic channel.
The spin polarizations lie in the $xy$-plane and the spiral
wave vector is along the $z$-axis.
The spiral wave vector $\vec q$ should be at the order of the inverse
of the system size in order to minimize the energy cost of spin twisting,
which may further depend on concrete boundary conditions.

An interesting point is that because of the coupling between
the ferromagnetic and ferro-nematic orders, the Fermi surfaces are
distorted \cite{fregoso2010}.
The oscillations of the distorted Fermi surfaces couple to spin
waves and bring Landau damping to spin waves.
This may result in non-Fermi liquid behavior to fermion excitations.
Similar effects in the nematic Fermi liquid state
have been extensively studied before in the literature
\cite{oganesyan2001,garst2010,metlitski2010}.

The next sub-leading instability is in the $J=1^-$ channel with $L=1$
and $S=1$, which has been studied in Ref. [\onlinecite{sogo2012,li2012a}].
For $J_z=0$, the generated SO coupling at the single
particle level exhibits the 3D Rashba-type as
\bea
H_{so,1^-}= |n_z|  \sum_k \psi^\dagger_\alpha(\vec k )
(k_x \sigma_y -k_y \sigma_x)_{\alpha\beta}
\psi_\beta(\vec k),
\label{eq:so}
\eea
where $|n_z|$ is the magnitude of the SO order parameter.
The magnetic dipolar interaction already possesses the SO nature.
The Pomeranchuk instability brings it to the single
particle level by breaking rotational symmetry and parity.

The instability of the $J=1^-$ sector is similar to but
fundamentally different from the Pomeranchuk instability
in the $L=S=1$ channel studied in Refs. [\onlinecite{wu2004,wu2007}].
Different from the magnetic dipolar systems, the starting Hamiltonian
in the latter case has no SO coupling at all.
After instability occurs, an effective SO coupling
appears at the single particle level.
In particular, a $\vec k \cdot \vec \sigma$ type SO coupling
can be generated with total angular momentum $J=0$.
It is generated from the SO decoupled
interactions through a phase transition, and thus was denoted as
spontaneous generation of SO coupling \cite{wu2004,wu2007}.
They are a particle-hole channel analogy of the $p$-wave
triplet Cooper pairings of the isotropic $^3$He-$B$ phase \cite{balian1963}
and the anisotropic $^3$He-$A$-phase \cite{anderson1961,brinkman1974},
respectively.

%-----------------------------------------------------------
\subsection{The SO coupled collective modes
-- the topological zero sound}
\label{sect:socoll}

We review the study of the collective modes in the SO coupled
Fermi liquid theory \cite{sogo2012,li2012a}.
The generalized Boltzmann equation including the spin degree of freedom
can be written as \cite{negele1988}
\bea
&&\frac{\partial}{\partial t} n(\vec r, \vec k, t)
-\frac{i}{\hbar} [\epsilon(\vec r, \vec k, t),  n(\vec r, \vec k, t)]
\nn \\
&+&\frac{1}{2}\sum_i
\Big\{ \frac{\partial \epsilon (\vec r, \vec k, t)}{\partial k_i},
\frac{\partial n (\vec r, \vec k, t)}{\partial r_i} \Big\} \nn \\
&-&
\frac{1}{2} \sum_i
\Big\{ \frac{\partial \epsilon (\vec r, \vec k, t)}{\partial r_i},
\frac{\partial n (\vec r, \vec k, t)}{\partial k_i} \Big\}
\label{eq:boltzmann}
=0,
\eea
where $n_{\alpha\alpha^\prime}(\vec r, \vec k, t)$ and  $\epsilon_{\alpha\alpha^\prime}
(\vec r, \vec k, t)$ are the density and energy matrices
for the coordinate $(\vec r, \vec k)$ in the phase space;
$[, ]$ and $\{, \}$ mean the commutator and
anti-commutator, respectively.
After linearizing Eq. \ref{eq:boltzmann} and expanding it
in the plane-wave bases
\bea
\delta n_{\alpha\alpha^\prime}(\vec r, \vec k)
=\sum_q \delta n_{\alpha\alpha^\prime}(\vec k) e^{i\vec q \cdot \vec r- i
\omega t},
\eea
we arrive at
\bea
\delta n_{\alpha\alpha^\prime} (\hat k)&-&
\frac{1}{2}\frac{\cos\theta_k}{s-\cos \theta_k} \sum_{\beta\beta^\prime}
\int d\Omega_{k^\prime}
\frac{N_0}{4\pi} f_{\alpha\alpha^\prime, \beta\beta^\prime}(\hat k, \hat k^\prime)
\nn \\
&\times& \delta n_{\beta\beta^\prime}(\hat k^\prime)=0,
\label{eq:so-zero}
\eea
where $s$ is the dimensionless parameter $\omega/(v_f q)$.
Without loss of generality, the propagation direction of
the wave vector $\vec q$ is defined as the $z$-direction.

By using the SO coupled bases $\delta n_{JJ_z;LS}$,
Eq. \ref{eq:so-zero} is rewritten as
\bea
\delta n_{JJ_z;LS}&+&\sum_{J^\prime;LL^\prime} K_{JJ_zLS;J^\prime J_z L^\prime S}(s)
F_{J^\prime J_zL^\prime S; J^\prime J_z L^{\prime\prime} S}\nn \\
&\times& \delta n_{J^\prime J_z L^{\prime\prime} S}=0,
\label{eq:sound_eign1}
\eea
The matrix kernel $K_{JJ_zLS;J^\prime J_z L^\prime S}$ reads
\bea
K_{JJ_zLS;J^\prime J_z L^\prime S} (s) &=&\sum_{m s_z}
\avg{LmSs_z|JJ_z} \avg{L^\prime m S s_z| J^\prime J_z} \nn \\
&\times& \Omega_{LL^\prime;m}(s),
\eea
where $\Omega_{LL^\prime}(s)$ is equivalent to the particle-hole channel
Fermi bubble in the diagrammatic method defined as
\bea
\Omega_{LL^\prime;m}(s)=-\int d\Omega_{\hat k} Y^*_{Lm}(\hat k)
Y_{L^\prime m }(\hat k) \frac{\cos \theta_k }{s-\cos\theta_k}.
\eea

%-------------------------------------------------------------------
\begin{figure}[tbp]
\centering\epsfig{file=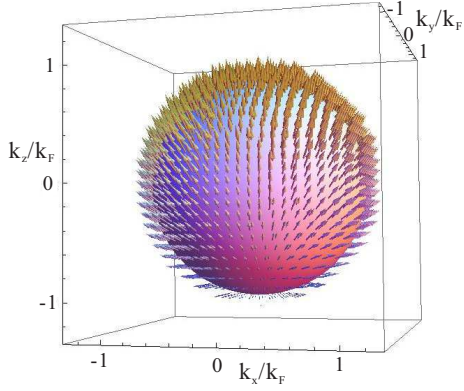,clip=1,width=0.7\linewidth, angle=0}
\caption{The spin configuration (Eq. \ref{eq:spinconfig}) of the zero
sound mode over the Fermi surface shows the hedgehog type topology.
Although the hedgehog configuration is distorted
in the $z$-component, its topology does not change for any values
of $\lambda_m$ describing the interaction strength.
From Ref. [\onlinecite{li2010}].
}
\label{fig:spinconfig}
\end{figure}
%------------------------------------------------------------------

The largest positive Landau parameter lies in the $(1^+001)$ channel,
which can support propagating modes.
Since $\vec q$ breaks parity and the 3D rotation symmetries,
the $(1^+001)$ channel couples to other channels with $J_z=0$.
The Landau parameters of orbital partial wave channels with $L\ge 2$
are small, which will be neglected henceforth.
There are three spin channel modes with $L=S=1$ and $J_z=0$
denoted as $(0^-011)$, $(1^-011)$, and $(2^-011)$.
Even in the presence of $\vec q$, the system still
possesses the reflection symmetry with respect to any plane
including $\vec q$.
By writing down their bilinear fermion expressions,
we can check that among the above modes,
$(1^+001)$, $(0^-011)$, $(2^-011)$ are odd, and $(1^-011)$ is
even under this reflection operation.
Furthermore, the $(2^-011)$ channel can be neglected because
the Landau parameter in this channel is about one order smaller
than those in $(1^+001)$ and $(0^-011)$.

Now, we only keep two coupled modes $(1^+001)$ and $(0^-011)$.
Using  the following relations
\bea
K_{1001;1001}(s)&=& \Omega_{00;0}(s),\nn \\
K_{1001;0011}(s)&=& K_{0011;1001}(s)= s \Omega_{00;0}(s), \nn \\
K_{0011,0011}(s)&=&\sum_m |\avg{1m1-m|00}|^2 \Omega_{11;m}(s)\nn \\
&=& \Omega_{00;0}(s),
\eea
we reduce the coupled $2\times 2$ matrix equation based on
Eq. \ref{eq:sound_eign1} into
\bea
\Omega_{00;0}(s)&=&
1-\frac{s}{2} \ln |\frac{1+s}{1-s}|+ i\frac{\pi}{2} s
\Theta(s<1) \nn \\
&=&
\frac{F_+ \pm \sqrt{F_+^2 +4(s^2-1)F_{\times}}}
{2(s^2-1)F_{\times}},
\label{eq:sound_velocity}
\eea
where
\bea
F_+=F_{1001:1001}+F_{0011;0011},
F_{\times}=F_{1001:1001} F_{0011;0011}. \nn
\eea

For the two branches of Eq. \ref{eq:sound_velocity}, only
the one with the minus sign possesses the solution with $s>1$,
as required by the condition of the undamped
collective mode of the Fermi liquid.
Since $n_{1^+001}(\vec q)=\sum_{\vec k} \psi^\dagger(\vec k+\vec q)
\sigma_z \psi_\beta(\vec k)$, and
$n_{0^-011}(\vec q)=\sum_{\vec k} \psi^\dagger(\vec k+\vec q)
(\vec k\cdot \vec \sigma )\psi_\beta(\vec k)$, the former
mode describes spin oscillation along the %direction of the
direction of $\vec q$, and the latter exhibits a
hedgehog configuration of spin distribution on the Fermi surface.
The eigen-mode is a hybridization between them, which can be represented as
\bea
\vec s(\vec r, \vec k, t) = \left(\begin{array}{c}
u_2 \sin\theta_{\vec k} \cos\phi_{\vec k} \\
u_2 \sin\theta_{\vec k} \sin\phi_{\vec k} \\
u_2 \cos \theta_{\vec k} +u_1
\end{array}
\right) e^{i(\vec q \cdot \vec r -s q v_f t)},
\label{eq:spinconfig}
\eea
where $(u_1, u_2)^T$ is the eigenvector for the collective mode.
For all the values of $\lambda_m$, $|u_2|>|u_1|$ is satisfied.
Thus, the spin configuration, as shown in Fig. \ref{fig:spinconfig},
is topologically non-trivial with the Pontryagin index $\pm 1$. %,
%whose sign periodically flips as varying time and spatial
%coordinates, {\it i.e.}, at the nodes of the sound wave.
%It can be considered as a topological zero sound.
The sign of the Pontryagin index periodically flips at the nodes of
the sound wave as the time and spatial
coordinates varies.
This collective mode can be considered as the topological zero sound.

%%%%%%%%%%%%%%%%%%%%%%%%%%%%%%%%%%%%%%%%%%%%%%%%%%%%%%%%%%%%
%%%%%%%%%%%%%%%%%%%%%%%%%%%%%%%%%%%%%%%%%%%%%%%%%%%%%%%%%%%%
%%%%%%%%%%%%%%%%%%%%%%%%%%%%%%%%%%%%%%%%%%%%%%%%%%%%%%%%%%%%
\section{Unconventional triplet Cooper pairing with
multi-component dipolar Fermi gases}
\label{sect:el_dp_pair}

The $p$-wave ($L=1$) triplet ($S=1$) Cooper pairing is a celebrated
unconventional pairing superfluid state which has been a research focus
of condensed matter physics for decades
(see Ref. [\onlinecite{leggett1975,volovik2009}] for reviews).
The typical system is the superfluid $^3$He which exhibits both the
isotropic $B$ phase \cite{balian1963} and the anisotropic $A$-phase
\cite{anderson1961,brinkman1974}.
So far, the most accepted pairing mechanism is spin fluctuations
arising from the prominent ferromagnetic tendency because
there exists a hard core part in the interaction between
two $^3$He atoms.

As explained in Sect. \ref{sect:intro}, the
anisotropic dipolar interaction provides a novel and robust
mechanism for the $p$-wave spin triplet Cooper pairing and
the competition between the singlet and triplet channel
pairing instabilities
\cite{you1999,baranov2002,baranov2004,
bruun2008,levinsen2011,potter2010,lutchyn2010,
zhao2010,wu2010,shi2009,qi2013,shi2013,han2010,liu2014}.
Furthermore, the coexistence of the singlet and triplet
pairing symmetries in the electric dipolar systems naturally
leads to a novel TR symmetry breaking mechanism first
pointed out in  Ref. [\onlinecite{wu2010}].
Recently, this mechanism is also studied in a wide context of
superconducting systems \cite{goswami2014,hinojosa2014}.
For example, it has been later proposed in the iron-based
superconductors \cite{hinojosa2014}, and its topological
electro-magnetic responses have been also studied \cite{goswami2014}.
In fact, the TR symmetry breaking pairing is also a
research focus in condensed matter physics, such as in the
study of high T$_c$
cuprates (e.g. YBa$_2$Cu$_3$O$_{6+x}$) \cite{xia2008} and ruthenates
(e.g. Sr$_2$RuO$_4$) \cite{xia2006}.

In this section, we will first present a brief overview %the notation
of the $p$-wave pairing in the superfluid $^3$He system in Sect.
\ref{sect:He3}.
The pairing symmetries in the single-component and two-component
electric dipolar fermion systems are reviewed in Sect.
\ref{sect:single} and Sect.
\ref{sect:two_comp_pair}, respectively.
The effect of the TR reversal symmetry breaking
is reviewed in Sect. \ref{sect:TR}.

%-----------------------------------------------------------
\subsection{A quick overview of the $p$-wave triplet pairing in the
superfluid $^3$He system}
\label{sect:He3}

The general structure of the $p$-wave spin
triplet pairing takes the form of a $2\times 2$ symmetric matrix
$\Delta_{\alpha\beta}(\vec k)$ in momentum space, which can be
conveniently represented by
the $d$-vector
as \cite{leggett1975}
\bea
\Delta_{\alpha\beta}(\vec k)= \Delta(\vec k) \hat d_\mu (\vec k)
(i\sigma^\mu \sigma^y)_{\alpha\beta},
\label{eq:pair}
\eea
where $\Delta(\vec k)$ is a complex number; $\hat d$ describes
the spin degree of freedom, which is a
complex unit 3-vector satisfying $\hat d^* \cdot \hat d=1$;
$i\sigma^\mu \sigma^y (\mu=x,y,z)$ form the bases of the $2\times 2$
symmetric matrices for the triplet pairing.
The so-called unitary pairing means that
$\Delta^\dagger(\vec k) \Delta(\vec k)$ equals
the identity matrix up to a constant, i.e.,
$\Delta(\vec k)$ is proportional to a unitary matrix.
In terms of the $d$-vector language, the unitary pairing
corresponds to $\hat d^* \times \hat d=0$, i.e., $\hat d$ is
equivalent to a unit real vector up to an overall
complex phase.

The two most prominent superfluid phases of $^3$He are $B$ and $A$ phases,
both of which belong to the class of unitary pairing.
In the $B$ phase \cite{balian1963}, a typical pairing matrix
structure is
\bea
\Delta^B_{\alpha\beta}(\vec k)=
\Delta \hat k \cdot (\vec \sigma i\sigma^y)_{\alpha\beta},
\eea
where $\Delta$ is a complex constant.
In this
configuration, $L$ and $S$ are combined into $J=0$, and
thus, the $B$ phase is rotationally invariant and also a fully gapped phase.
In other words, the $B$ phase spontaneously breaks the
relative SO symmetry, exhibiting the SO coupled paired structure.
It maintains TR and parity symmetries.
Furthermore, it has also been recently found that
the $B$ phase is actually a 3D topological superfluid phase.
In comparison, the $^3$He-$A$ phase is anisotropic, in which $L$ and
$S$ are decoupled and $J$ is not well-defined
\cite{brinkman1974,anderson1961}.
A typical pairing matrix for the $A$ phase is
\bea
\Delta^A_{\alpha\beta}(\vec k)=
\Delta (\hat k_x+i\hat k_y) (\hat d \cdot \vec \sigma i\sigma^y)_{\alpha\beta},
\eea
in which $\hat d$ is an arbitrary unit 3-vector.
It breaks TR symmetry and exhibits nodal points
on the north and south poles on the Fermi surface.

We will see in this section and in Sect. \ref{sect:mg_dp_pair}
that both the electric and magnetic dipolar
fermion systems support novel $p$-wave triplet structures
which are different from the $^3$He-$B$ and $A$ phases.

%---------------------------------------------------------------
\subsection{The $p_z$-wave pairing with the
single-component dipolar fermions}
\label{sect:single}

%-------------------------------------------------------------------
\begin{figure}[tbp]
\centering\epsfig{file=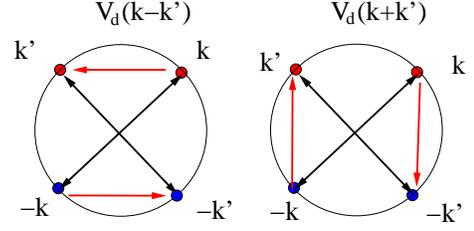,clip=1,width=0.7\linewidth, angle=0}
\caption{Pairing interaction matrix elements for the single
component dipolar interaction $V_t(\vec k;\vec k^\prime)=
V_d(\vec k -\vec k^\prime)-V_d(\vec k +\vec k^\prime)$ due to the
two-particle interference and the Fermi statistics.
}
\label{fig:pair_int}
\end{figure}
%---------------------------------------------------------------

The pairing symmetry structure of the single-component dipolar system
was studied in early works of Refs. [\onlinecite{you1999,baranov2002}].
In this case, the gap function is simplified as a complex number,
and thus, the $d$-vector notation is not needed.

In real space, this $p_z$-pairing symmetry is also clear since
the electric dipolar interaction is the most attractive if the displacement
vector between two fermions is along the $z$-axis.
Below, we present the partial wave analysis in momentum space.
The pairing interaction can be expressed at the
mean-field level as \cite{you1999,baranov2002}
\bea
H_{pair}&=& \frac{1}{2V_0}
\sum_{k,k^\prime} V_d(\vec k- \vec k^\prime)
\psi^\dagger(\vec k) \psi^\dagger(-\vec k) \psi(-\vec k^\prime)
\psi(\vec k^\prime),
\nn \\
&=&\frac{1}{4V_0}\sum_{k,k^\prime} V_t(\vec k; \vec k^\prime)
\psi^\dagger(\vec k) \psi^\dagger(-\vec k) \psi(-\vec k^\prime)
\psi(\vec k^\prime),  \ \ \, \ \ \, \ \ \,
\eea
where
\bea
V_t(\vec k;\vec k^\prime)=V_d(\vec k-\vec k^\prime)
-V_d(\vec k +\vec k^\prime)
\eea
satisfying $V_t(\vec k;\vec k^\prime)
=V_t(-\vec k;\vec k^\prime)=-V_t(\vec k, \vec k^\prime)$
as shown in Fig. \ref{fig:pair_int}.
The symbol $V_t$ is used because the pairing analysis here
is the same as that for the
triplet pairing of the two-component dipolar fermions
in Sect. \ref{sect:two_comp_pair}.

The BCS mean-field gap equation is
\bea
\Delta(\vec k)&=&-\int \frac{d^3k^\prime}{(2\pi)^3}
V_{t}(\vec k;\vec k^\prime)
[K(\vec k^\prime)-\frac{1}{2\epsilon_k}]
\Delta(\vec k^\prime), \ \ \,
\label{eq:gap_sing}
\eea
where $E(\vec k)=\sqrt{(\epsilon_k-\mu)^2+\Delta^2 (\vec k)}$
and $K(\vec k)=\tanh[\frac{\beta}{2}E(\vec k)]/[2E(\vec k)]$.
The integral of Eq. \ref{eq:gap_sing} has been regularized
following the $t$-matrix method at the level of
the Born approximation. % level.
This regularization is
equivalent to truncate the energy away from the Fermi energy
$\pm\bar\omega$, and $\bar\omega$ is at the order of the
Fermi energy \cite{baranov2002}.

\begin{figure}[tbp]
\centering\epsfig{file=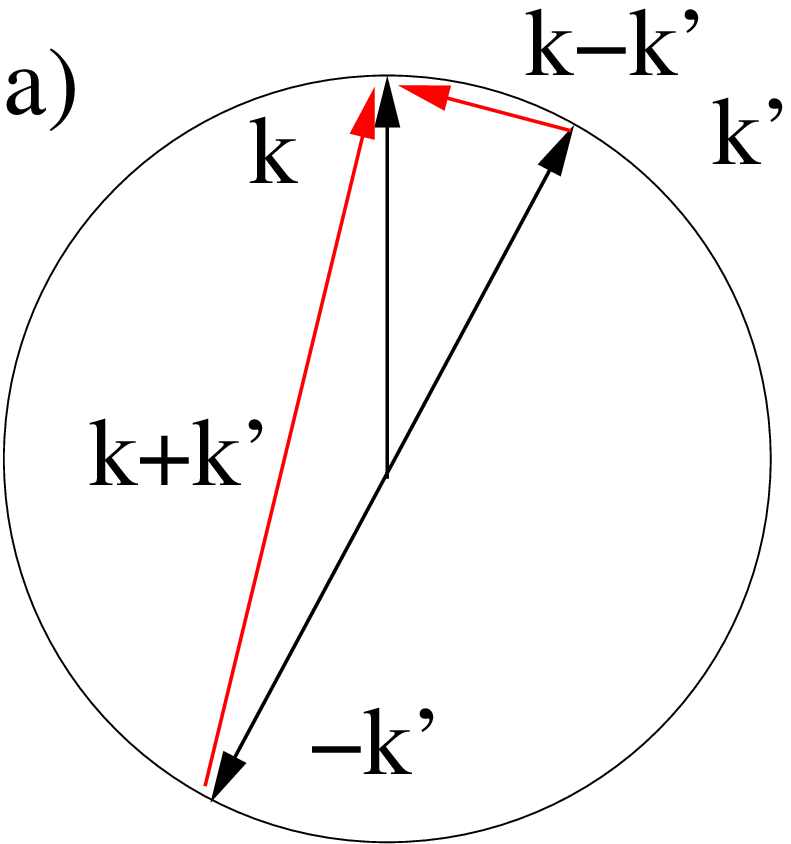,clip=1,width=0.3\linewidth, angle=0}
\hspace{4mm}
\centering\epsfig{file=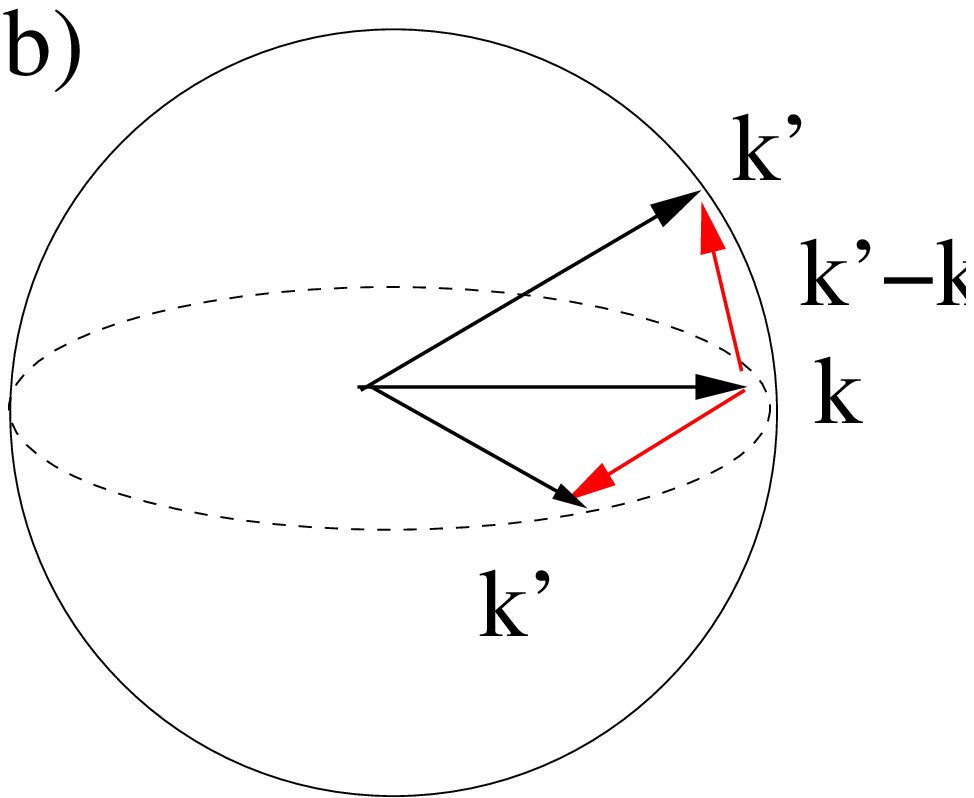,clip=1,width=0.36\linewidth, angle=0}
\caption{Pairing interaction matrix elements for the dipolar interaction
$V_t(\vec k;\vec k^\prime)$.
$a$) For the case of $\vec k \parallel \hat z$, and $\vec k^\prime
\rightarrow \vec k$, $V_t(\vec k;\vec k^\prime)=V_t(\vec k;-\vec k^\prime)<0$.
$b$) For the case of $\vec k \parallel \hat x$, and $\vec k^\prime
\rightarrow \vec k$, $V_t(\vec k;\vec k^\prime)$ varies and its
angular average is positive.
}
\label{fig:partial_wave}
\end{figure}

Although the $p_z$-pairing is very intuitive in the real space,
it is not so obvious in momentum space.
Before performing the partial-wave analysis, a qualitative
momentum space picture can reveal why the $p_z$-pairing
is natural as shown in Fig. \ref{fig:partial_wave} (a).
Let us set $\vec k\parallel \hat z$ and $\vec k^\prime
\rightarrow \vec k$, then $(\vec k -\vec k^\prime) \perp \hat z$ and
$(\vec k +\vec k^\prime) \parallel \hat z$, thus
\bea
V_t(\vec k; \vec k^\prime)=-4\pi d^2<0, \ \ \,
V_t(\vec k;-\vec k^\prime)= 4\pi d^2>0,
\eea
which favors the pairing in polar regions with an odd parity.
On the other hand, if we set $\vec k$ in the equatorial plane,
say, $\vec k \parallel \hat x$, and also $\vec k^\prime
\rightarrow \vec k$, then $\vec k-\vec k^\prime$ lies in the
$yz$-plane.
The value of $V_d(\vec k -\vec k ^\prime)$ depends on the polar
angle of the vector of $\vec k-\vec k^\prime$.
Its average is $\frac{2\pi}{3} d^2$
and that of $V_d(\vec k +\vec k^\prime)=-\frac{4\pi}{3}d^2$,
thus the average  of $V_t(\vec k;\vec k^\prime)$ as
$\vec k^\prime \rightarrow \vec k$ for $\vec k$ in the
equatorial plane is positive.
This means that the pairing amplitude is suppressed in the equatorial plane.
Combining the pairing structures both in the polar region and the
equatorial plane,
it is clear that the pairing symmetry is mostly of the $p_z$-wave,
which is also consistent with the real space picture of pairing.

%---------------------------------------------------------
%------------------------------------------------------------------

Around $T_c$, the gap equation can be linearized.
The standard pairing eigenvalue analysis is performed
by defining the eigenvectors $\phi_{t,s}^i (\hat k)$
satisfying the eigen-equation
\bea
N_0 \int d\Omega_{k^\prime} V_{t} (\hat k; \hat k^\prime)
\phi^i_{t} (\hat k^\prime)=w_{t}^i \phi_{t,s}^i(\hat k),
\label{eq:pair_part_wv}
\eea
where $w_{t}^i$ are eigenvalues; $N_0$ is the density of
states on the Fermi surface; $i$ is the index of eigenvectors.
The actual pairing occurs in the channel with the lowest negative
eigenvalue.
The spherical harmonics decomposition of  $V_{t,s} (\vec k;\vec k^\prime)$
reads
\bea
\frac{N_0}{4\pi}V_{t}(\vec k;\vec k^\prime)=\sum_{l,l^\prime;m}
V_{ll^\prime;m}
Y_{lm}^*(\Omega_k) Y_{l^\prime m} (\Omega_{\vec k^\prime}),
\label{eq:pair_decomp}
\eea
where $l$ and $l^\prime$ only take odd values.
Compared with Eq.\ref{eq:landau_inter} and Eq. \ref{eq:landauhmncs},
we arrive at
\bea
V_{ll^\prime;m}=-\frac{2N_0 f_{ll^\prime;m}}{\sqrt{(2l+1)(2l^\prime+1)}}.
\label{eq:pair_matrix}
\eea

The diagonalization shows that the most negative eigenvalue lies
in the sector with $m=0$, whose eigenvalue and eigenvector are
\bea
w^{i=0}_{t,m=0}=-3.82\lambda, \ \ \,
\phi^z(\Omega_k)\approx 0.99 Y_{10}-0.12 Y_{30}.
\label{eq:pz_pair}
\eea
The above pairing eigenvector shows that the pairing symmetry
is mainly of the $p_z$-type, in agreement with the  intuitive
real space analysis.
The pairing eigen-values in the sector of $m=0$ can even been
solved analytically as shown in Ref. \onlinecite{baranov2002} as
$w_{t,m=0}^i=-12\lambda/[\pi(2i+1)^2]$.
The gap function of Eq. \ref{eq:pz_pair}
vanishes for quasi-particle momenta lying in the equatorial plane,
and thus exhibits a nodal plane in the Bogoliubov excitation spectra.

As shown in Ref. [\onlinecite{baranov2002}], the standard
mean-field value of $T_c$ is related to the eigenvalue
solved above as
\bea
T_c=\frac{2e^\gamma \bar \omega}{\pi} e^{\frac{1}{w_t^0}},
\label{eq:Tc}
\eea
where $\gamma=0.5772$ is the Euler constant.
In Ref.  [\onlinecite{baranov2002}], a further analysis
based on Gor'kov, Melik-Barkhudarov (GM) approach \cite{gorkov1961}
is performed, which takes into account the media polarization effect
due to the virtual process of creating particle-hole excitations.
This GM approach shows that the energy cutoff effect can
be approximated by $\bar\omega\approx 0.42 \epsilon_f$.

%------------------------------------------------------------
\subsection{Competition between the triplet and singlet
pairings in the two-component dipolar Fermi gases}
\label{sect:two_comp_pair}

The new ingredient of the two-component dipolar fermions is that both
the spin singlet and triplet pairings are allowed
\cite{wu2010,shi2009,qi2013,shi2013}.
We define the pairing operators in these two sectors as
\bea
P_{s}(\vec k)&=&\frac{1}{\sqrt 2} \tr \Big [P(\vec k) (-i\sigma^y)\Big ],
\nn \\
P_{t}^\mu(\vec k)&=& \frac{1}{\sqrt 2} \tr \Big[P(\vec k) (-i\sigma^y
\sigma^\mu) \Big ],
\eea
with $\mu=x,y,z$, respectively,
where $P_{\alpha\beta}(\vec k)=\psi_\alpha(\vec k) \psi_\beta(-\vec k)$.
Then the pairing Hamiltonian is expressed as
\bea
H_{pair}
&=& \frac{1}{2V_0} \sum_{k,k^\prime,} \Big\{ ~V_t(\vec k;\vec k^\prime)
~[~\sum_{\mu=x,y,z} P^{\dagger,\mu}_t(\vec k) P_t^\mu(\vec k^\prime)~]\nn \\
&+& V_s(\vec k;\vec k^\prime) P^\dagger_s(\vec k) P_s(\vec k^\prime)~\Big\},
\label{Eq:polarpair}
\eea
where $V_{t,s}(\vec k;\vec k^\prime)=V_d(\vec k-\vec k^\prime) \mp
V_d(\vec k +\vec k^\prime)$ are pairing interactions in the
singlet and triplet channels, respectively.
The Bogoliubov quasiparticle spectra become
$E_i(\vec k)=\sqrt{(\epsilon_k-\mu)^2+\lambda^2_i(\vec k) }$,
and $\lambda^2_{1,2}(\vec k)$ are the eigenvalues of the
positive-definite Hermitian matrix $\Delta^\dagger(\vec k)
\Delta(\vec k)$.

The gap equation takes the matrix form as
\bea
\Delta_{\alpha\beta}(\vec k)=-\int \frac{d^3k^\prime}{(2\pi)^3}
V_d(\vec k-\vec k^\prime) \avg{|\psi_\alpha(\vec k) \psi_\beta(-\vec k)|},
\eea
where $\avg{||}$ means the thermal ensemble average.
$\Delta_{\alpha\beta}(\vec k)$ can be decomposed into
the singlet and triplet channel pairings as
\bea
\Delta_{\alpha\beta}(\vec k)=\Delta_{s} (\vec k) i\sigma^y_{\alpha\beta}
+\Delta_{t,\mu} (\vec k) (i\sigma^\mu \sigma^y)_{\alpha\beta},
\eea
in which $\Delta_s$ and $\Delta_{t,\mu}$ satisfy
\bea
\Delta_{t(s),\mu}(\vec k)&=& -\frac{1}{2}\sum_{i}\int \frac{d^3k^\prime}{(2\pi)^3}
V_{t(s)}(\vec k;\vec k^\prime)
[K_i(\vec k^\prime)\nn \\
&-&\frac{1}{2\epsilon_k}]
\Delta_{t(s),\mu} (\vec k^\prime),
\label{eq:gap}
\eea
where $K_i(\vec k)=\tanh[\frac{\beta}{2}E_i(\vec k)]/[2E_i(\vec k)]$.
After linearizing the gap equation around $T_c$, we perform the
eigenvalue analysis for the pairing problem.
The eigen-equation of the triplet sector is the same as
Eq. \ref{eq:pair_part_wv}, while that of the singlet
sector can be obtained by replacing $V_t(\vec k;\vec k^\prime)$
with $V_s(\vec k; \vec k^\prime)$.
The spherical harmonics decomposition of $V_s(\vec k; \vec k^\prime)$
can be done in a similar way to  Eq. \ref{eq:pair_decomp},
and the resultant $V_{ll^\prime;m}$ takes the same form as
Eq. \ref{eq:pair_matrix} but with $l, l^\prime$ only
taking even integer values.
However, the Hartree interaction does not exist in the pairing channel,
and thus, for $l=l^\prime=m=0$, $V_{00;0}=0$ for the case of the purely
dipolar interaction.

The analysis of pairing eigenvalues in the triplet sector is the same as
that in the single component case.
Thus, the leading pairing symmetry still lies in the $p_z$-channel.
The pairing eigenvalue and eigenvector are the same as
Eq. \ref{eq:pz_pair}, and $T_c$ is still approximately determined
by Eq. \ref{eq:Tc}.
We can express the triplet pairing in terms of the $d$-vector as
\bea
\Delta_{t,\mu}(\vec k)=\Delta_{t} e^{i\gamma} \phi^z(\Omega_k) \hat d_\mu,
\eea
where $\gamma$ is the U(1) phase and $\Delta_t$ is the pairing amplitude.
This spin-triplet pairing breaks the U(1) gauge symmetry in the
charge channel and the SU(2) symmetry in the spin channel, and thus,
there are two different low energy excitations: phonon and spin-wave modes.
This pairing is still invariant under a combined $Z_2$-symmetry
of $\hat d_\mu\rightarrow -\hat d_\mu$ and
$\gamma\rightarrow \gamma+\pi$ \cite{zhou2003,salomaa1987},
and thus, it supports two different classes of vortices: the
usual integer
vortex of superfluidity, and the half-integer quantum vortex of
superfluidity combined with a $\pi$-disclination of the $d$-vector.

In the singlet channel, the lowest eigenvalue and the
corresponding eigenvector are
\bea
w_{s}^0=-1.93\lambda, ~~~\psi^{s+d}(\Omega_k)\approx 0.6 Y_{00}-0.8 Y_{20},
\eea
respectively.
The eigenvector mixes the $s$ and $d_{k^2-3k_z^2}$-channels.
All other negative eigenvalues are small and negligible.
Although for the purely dipolar interaction, the singlet channel
pairing instability is significantly weaker than that of the
triplet channel.
Nevertheless, the matrix element $V_{00;0}$ may receive additional
contributions from the short-range $s$-wave scattering interaction,
which in principle is tunable through Feshbach resonances.
At $V_{00;0}/\lambda\approx -3.15$, the singlet and triplet
channel instabilities become degenerate.

Considering the competition between the spin triplet and singlet
channel pairings, we generally expect two pairing superfluid
transitions.
The first transition is triggered by the stronger pairing channel,
say, the triplet $p_z$-channel, which determines $T_{c1}$.
In the case that
the singlet channel pairing is weaker but nearly degenerate with
the triplet one, a second transition may occur at $T_{c2}$ as
further lowering temperature.
The mixing between the single and triplet pairings breaks parity,
and thus, the second transition is also a genuine phase transition.
The coupling between these two pairing channels can be
captured by the following Ginzburg-Landau (GL) free energy as
\cite{wu2010}
\bea
\Delta F&=& \gamma_1 (\vec \Delta_{t}^*\cdot \vec \Delta_t)
|\Delta_s|^2 + \gamma_2 \{\vec \Delta_{t}^*
\cdot \vec \Delta_{t}^* \Delta_s \Delta_s +cc\} \nn \\
&+&\gamma_3 |\vec \Delta_{t}^* \times \vec \Delta_{t}|^2,
\eea
where $\vec \Delta_t$ is a compact notation for $\Delta_{t,\mu} (\mu=x,y,z)$;
other non-gradient terms in the GL free energy only depend on
the magnitudes of order parameters.
$\gamma_3$ should be positive, such that $\vec \Delta_t$
can be described by a real $d$-vector multiplied by
a $U(1)$ phase.
The sign of $\gamma_2$  determines the relative phase between
$\Delta_s$ and $\vec \Delta_t$:
if $\gamma_2<0$, the phase difference between
$\Delta_s$ and $\vec \Delta_t$ is 0 or $\pi$;
if $\gamma_2>0$, the phase difference is $\pm\frac{\pi}{2}$.
The latter case gives rise to a novel pairing with TR
symmetry breaking as explained in Sect. \ref{sect:TR}.

In SO coupled systems, the coupling between the singlet and triplet
pairings through a spatial gradient is considered in Ref.
[\onlinecite{samokhin2006}], which leads to a spatially non-uniform state.
Nevertheless, because of the spin conservation, such a term is not allowed
in electric dipolar systems.

%-----------------------------------------------------
\subsection{TR symmetry breaking mixing
between the singlet and triplet  pairings}
\label{sect:TR}

When the singlet and triplet channel pairings coexist, %s,
 a natural
question is: what is the relative phase between these two pairing channels? %.
It was found in Ref. [\onlinecite{wu2010}] that, at
the mean-field level, the phase difference of $\pm\frac{\pi}{2}$
between $\Delta^s$ and $\Delta^{t,\mu}$ is favored, so %such
that
TR symmetry is spontaneously broken.
This is a general mechanism leading to the TR symmetry
breaking pairing, which applies for the mixing problem
between singlet and triplet pairings in the absence of SO
coupling in the weak coupling limit.
It has also been found later in Ref. [\onlinecite{qi2013,shi2013}] that the
TR symmetry breaking also exists in the resonance
interaction regime of the dipolar Fermi gases.

The reason for the above TR symmetry breaking effect is
that the weak coupling theory favors the unitary
pairing, {\it i.e.}, $\Delta^\dagger (\vec k) \Delta(\vec k)$ is
an identity matrix up to a factor.
A simple calculation shows that $\Delta^\dagger(\vec k)\Delta (\vec k)=
|\Delta^s|^2+|\Delta^{t,\mu}|^2 +\Re (\Delta^{s,*} \Delta^{t,\mu})
(\sigma^\mu)^T$.
The unitary pairing is achieved if and only if a phase difference
$e^{i\phi}=\pm i$ exists between $\Delta^s$ and $\Delta^{t,\mu}$.
More precisely, it can be proved following the method presented
in Ref. [\onlinecite{cheng2010}].
The quasi-particle spectra read $E_i=\sqrt{\xi^2+\lambda_i^2}
(i=1,2)$
with
\bea
\lambda_{1,2}^2&=&|\Delta_s \phi^{s+d}(\Omega_k)|^2
+|\Delta_t \phi^z(\Omega_k)|^2 \nn \\
 &\pm& 2\Re(\Delta^{*}_s \Delta_t)
\phi^{s+d}(\Omega_k) \phi^z(\Omega_k).
\eea
The quasiparticle contribution to the free energy is
$f(x)=-\frac{2}{\beta}\ln [2 \cosh \frac{\beta}{2}
\sqrt{\xi^2_k +x }]$, which satisfies $\frac{d^2}{dx^2} f(x) >0$
\cite{cheng2010}. %, t
Thus,
\bea
f(\lambda_1^2)+f(\lambda_2^2) \ge 2
f(\frac{\lambda_1^2+\lambda_2^2}{2})
\eea
and the minimum is reached at $\lambda_1^2=\lambda_2^2$.
This is precisely the condition of the unitary pairing.

Here is another intuitive view to see why the phase
difference $e^{i\phi}=\pm i$ is favored.
Without loss of generality, let us set $\hat d\parallel \hat z$,
and assume a relative phase difference $\phi$.
The ratio between %ration of
the pairing amplitudes is
\bea
\frac{\avg{\psi_\uparrow(\vec k )\psi_\downarrow(-\vec k)}}
{\avg{\psi_\downarrow(\vec k )\psi_\uparrow(-\vec k)}}
= \frac{\Delta_s + e^{i\phi}\Delta_t}{-\Delta_s+e^{i\phi} \Delta_t}.
\eea
Only when $e^{i\phi}=\pm i$, the magnitude of this ratio is 1,
such that the pairing strengths are the same.
Otherwise, say, if $e^{i\phi}=\pm 1$, the pairing amplitudes
of $\avg{\psi_\uparrow(\vec k )\psi_\downarrow(-\vec k)}$
and $\avg{\psi_\downarrow(\vec k )\psi_\uparrow(-\vec k)}$
are not equal. % non-equal.
Again, the pairing structure, whose gap magnitudes distribute over
the Fermi surface in a more uniform way, is usually favored.

%%%%%%%%%%%%%%%%%%%%%%%%%%%%%%%%%%%%%%%%%%%%%%%%%%%%%%%%%%%%%%%%%
\section{The $J$-triplet Cooper pairing with the magnetic dipolar
interaction}
\label{sect:mg_dp_pair}

The magnetic dipolar interaction also gives rise to novel Cooper pairing
structures possessing the SO coupled nature \cite{li2012}.
Again, below, we use the simplest case of $F=\frac{1}{2}$ as a prototype
model to explore the exotic pairing structure of %due to
the magnetic
dipolar interaction, which provides a novel and robust mechanism
for the $p$-wave ($L=1$) spin triplet ($S=1$) Cooper pairing.
It turns out that its pairing symmetry structure is markedly
different from those in the celebrated $p$-wave $^3$He $A$ and $B$
phases: the orbital angular momentum $L$ and spin $S$ of a Cooper
pair are coupled %entangled
into the total angular momentum  $J = 1$,
and thus dubbed as the $J$-triplet pairing.
In comparison, the $^3$He $B$ phase is isotropic in which
$J=0$; while, the $A$ phase is anisotropic in which $J$
is not well-defined \cite{leggett1975}.

Even within
the $J$-triplet sector,  there are still competing
instabilities regarding to different possibilities of $J_z$'s:
the helical polar state ($J_z=0$) which maintains TR
symmetry, and the axial state ($J_z=\pm 1$) which breaks TR symmetry.
The helical polar state exhibits a pair of nodes around which
the quasi-particle wavefunction exhibits winding numbers $\pm 1$,
thus it is a gapless Dirac pairing.
It is a TR invariant generalization of the $^3$He-$A$ phase
with coupled %entangled
spin and orbital degrees of freedom.
This %Such a
state was also proposed before, in the context of
superfluid $^3$He  as an intermediate
phase between the $^3$He-$B$ phase and the normal state \cite{fishman1987}.
In contrast, the axial pairing state exhibits a Weyl
type node around which the winding number is
$2$, and its %the
low energy spectrum is quadratic.

In this section, we review the SO coupled pairing structure of the
magnetic dipolar fermions.
An intuitive real space picture for the $J$-triplet pairing is presented
in Sect. \ref{sect:real}.
The partial-wave analysis for the pairing eigenvalues is performed
in Sect. \ref{sect:mgdp_partial}.
The structure of the nodal Bogoliubov quasi-particles is
given in Sect. \ref{sect:mgdp_topo}.

%-----------------------------------------
\subsection{The real space picture for the $J$-triplet pairing}
\label{sect:real}

%-------------------------------------------------------------------
\begin{figure}
\centering\epsfig{file=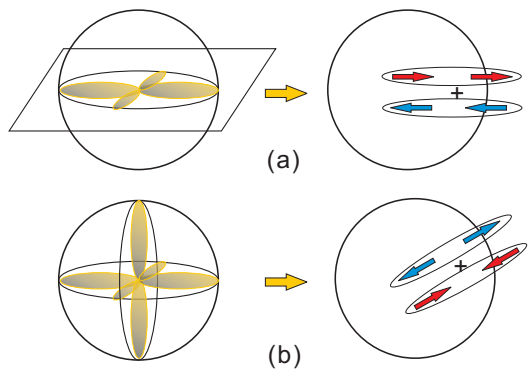,clip=1,width=0.8\linewidth, angle=0}
\caption{The spin configurations of the two-body states
with $a$) $J$=1 and $J_z=0$ and $b$) $J=J_z=0$.
The average magnetic dipolar interaction is attractive in a)
but repulsive in b). From Ref. [\onlinecite{li2012}].
}
\label{fig:twbdy}
\end{figure}
%------------------------------------------------------------------

We perform an intuitive real space analysis based on a simple two-body problem.
It can be shown that the magnetic dipolar interaction between
two spin-$\frac{1}{2}$ fermions vanishes in the total spin singlet
channel, and it only exists in the total spin triplet channel.
Naturally, the magnetic dipolar interaction leads to the triplet pairing.
Let us introduce a characteristic length scale at which the kinetic
energy equals the interaction energy at $a_{dp}=m g^2_F\mu_B^2/\hbar^2$
where $m$ is the fermion mass.
Since our purpose is to seek the most attractive angular partial-wave
channel,  without loss of generality, we can fix the inter-particle
distance at $a_{dp}$.
In the spin triplet channel, the lowest value of orbital angular
momentum is $p$-wave, and thus there are $3\times 3=9$ states
with $L=S=1$.
According to the total angular momentum $J$, they can be classified
into $J=0,1$ and $2$.
The interaction energies can be easily diagonalized in each sector as
\bea
E_0=E_{dp}, \ \ \, E_1=-\frac{1}{2} E_{dp}, \ \ \,  E_2=\frac{1}{10} E_{dp},
\eea
respectively, where $E_{dp}=g_F^2\mu_B^2/a^3_{dp}$.
Only the channel with $J=1$ can support bound states,
which is also confirmed from the momentum
space analysis below in Sect. \ref{sect:mgdp_partial}.

The reason why the $J$-triplet channel is the most attractive one
is because of its particular pairing spin configuration, which shows
a dominant ``head-to-tail'' configuration and thus the interaction
is attractive.
More precisely, let us denote the spin wavefunctions and the relative
orbital wavefunctions as $\chi_\mu$ and
$p_\mu(\hat\Omega)$, which satisfy
\bea
\{ \hat e_\mu \cdot (\vec S_1 +\vec S_2)\}  \chi_\mu =0, \ \ \,
(\hat e_\mu \cdot \vec L) p_\mu(\hat\Omega)=0,
\eea
for $\mu=x,y$, and $z$.
In other words, $\chi_\mu$ and $p_\mu(\hat\Omega)$ are polar eigenstates
of total spin and the relative orbital angular momentum, respectively.
In the sector of $J=1$, we define the SO coupled polar
state satisfying
\bea
(\hat e_\mu \cdot \vec J) \phi_\mu=0,
\eea
which can be expressed as $\phi_\mu(\Omega)=\frac{1}{\sqrt{2}}
\epsilon_{\mu\nu\lambda} \chi_\nu p_\lambda(\Omega)$.
In particular, for the state with $J_z=0$, it can be further expressed
as
\bea
\phi_z(\hat \Omega)
&=&\sqrt{\frac{3}{2}}\sin\theta \big\{|\alpha_{\hat e_\rho} \rangle_1
|\alpha_{\hat e_\rho} \rangle_2+|\beta_{\hat e_\rho} \rangle_1
|\beta_{\hat e_\rho}\rangle_2 \big\}, \ \ \,
\eea
where, $\hat e_\rho=\hat x \cos\phi+\hat y \sin \phi $; %and
 $|\alpha_{e_\rho}\rangle$ and $|\beta_{e_\rho}\rangle$ are eigenstates of
$\hat e_\rho \cdot \vec \sigma $ with eigenvalues of $\pm 1$,
respectively.
This shows the ``head-to-tail'' configuration in Fig. \ref{fig:twbdy} (a), and thus,
the corresponding interaction is attractive.

In contrast, the eigenstate of % with
$J=0$ shows the ``head-to-head''
configuration as shown in Fig. \ref{fig:twbdy} ($b$), and thus
the interaction in such a state is repulsive.
It  is expressed as
$\phi_0(\Omega)
=\frac{1}{\sqrt 2} \big\{|\alpha_\Omega \rangle_1 |\beta_\Omega\rangle_2+
|\beta_\Omega \rangle_1 |\alpha_\Omega\rangle_2 \big\},$
where $|\alpha_\Omega\rangle$ and $|\beta_\Omega\rangle$ are eigenstates
of $\hat \Omega \cdot \vec \sigma$ with the eigenvalues $\pm 1$,
respectively.

%--------------------------------------------------------------------
\subsection{The momentum space partial-wave analysis}
\label{sect:mgdp_partial}

In this part, we review the partial-wave analysis in momentum
space, which arrives at the same pairing symmetry as that obtained
through the real space analysis.

After the mean-field decomposition, the pairing Hamiltonian
of the magnetic dipolar system becomes
\bea
H_{mf}&=&\frac{1}{4}\int \frac{d^3k}{(2\pi)^3}
\Psi^\dagger (\vec k)
\left(\begin{array}{cc}
\xi(\vec k) I & \Delta_{\alpha\beta} (\vec k) \\
\Delta^*_{\beta\alpha}(\vec k) &-\xi(\vec k) I
\end{array}
\right) \Psi(\vec k), \nn \\
\label{eq:mf}
\eea
where
$\Psi(\vec k)=(\psi_\uparrow (\vec k), \psi_\downarrow(\vec k),
\psi_\uparrow^\dagger(-\vec k), \psi_\downarrow^\dagger(-\vec k) )^T$;
$\xi (\vec k)=\epsilon (\vec k) -\mu$.
The Bogoliubov quasiparticle spectra become
$E_i(\vec k)=\sqrt{\xi_k^2+\lambda^2_i(\vec k) }$
where $i=1,2$ and
$\lambda^2_i(\vec k)$ are the eigenvalues of the positive-definite
Hermitian matrix defined as $\Delta^\dagger(\vec k) \Delta(\vec k)$.

The pairing matrix $\Delta_{\alpha\beta}$ is defined as
\bea
\Delta_{\alpha\beta}=\sum_{S_z}\langle 1S_z
|\frac{1}{2}\alpha\frac{1}{2}\beta\rangle^*\Delta_{S_z},
\eea
where $\langle 1S_z|\frac{1}{2}\alpha\frac{1}{2} \beta\rangle$
is the Clebsch-Gordan coefficient for two spin-$\frac{1}{2}$
states to form the spin triplet.
$\Delta_{S_z}$ satisfies the mean-field gap function as
\bea
\Delta_{S_z}(\vec k)
&=&-\frac{1}{2}\sum_i \int \frac{d^3 k^\prime}{(2\pi)^3}
V_{S_zS_z^\prime}(\vec k;\vec k^\prime)
[K_i(\vec k^\prime)-\frac{1}{2\epsilon_k}]
\nn \\
&\times&
\Delta_{S^\prime_z} (\vec k^\prime),
\label{eq:gap_2}
\eea
where $K_i(\vec k^\prime)=\tanh[\frac{\beta}{2}
E_{i}(\vec k^\prime)]/[2E_i (\vec k^\prime)]$, and
the integral in Eq. \ref{eq:gap_2} is already normalized
following the standard procedure \cite{baranov2002}.

The interaction matrix element in Eq. \ref{eq:gap_2} is defined as
\bea
V_{S_zS_z^\prime}(\vec k;\vec k^\prime)&=&
\frac{1}{2}\sum_{\alpha\beta\beta^\prime\alpha^\prime}
\langle 1S_z|\frac{1}{2} \alpha \frac{1}{2}\beta\rangle
\langle 1S_z^\prime|\frac{1}{2} \alpha^\prime \frac{1}{2}\beta^\prime\rangle^*
\nn \\
&\times& \big\{ V_{\alpha\beta,\beta^\prime\alpha^\prime}(\vec k-\vec k^\prime)
-V_{\alpha\beta,\beta^\prime\alpha^\prime}(\vec k+\vec k^\prime)\big\}. \nn
\eea
The spherical harmonics decomposition of  $V_{S_zS_z^\prime}
(\vec k;\vec k^\prime)$ can be formulated as
\bea
\frac{N_0}{4\pi} V_{S_zS_z^\prime}(\vec k;\vec k^\prime)
&=&\sum_{LM,L^\prime M^\prime}
V_{LMS_z;L^\prime M^\prime S_z^\prime}
Y_{LM}^*(\Omega_k)\nn \\
&\times&
 Y_{L^\prime M^\prime} (\Omega_{\vec k^\prime}),
\label{eq:partialwave}
\eea
where $L=L^\prime$, or, $L=L^\prime\pm2$, and $L,L^\prime$ are odd integers.
The expressions of the dimensionless matrix elements
$V_{LMS_z;L^\prime M^\prime S_z^\prime}$ are the same as those
in Eq. \ref{eq:mgdr_landau} except an overall minus sign and
a numeric factor.

The free energy can be calculated as
\bea
F&=&-\frac{2}{\beta}\sum_{i=1,2} \int \frac{d^3k}{(2\pi)^3}\ln \big[2\cosh
\frac{\beta E_{\vec k,i} }{2}\big] \nn \\
&-&\frac{1}{2}   \sum_{S_z,S_z^\prime}
\int \int \frac{d^3k}{(2\pi)^3} \frac{d^3k^\prime}{(2\pi)^3}
\big \{ \Delta^*_{S_z}(\vec k) V^{-1}_{S_z S_z^\prime}(\vec k; \vec k^\prime)
\nn \\
&\times&
\Delta_{S_z^\prime}(\vec k)
\big\},
\label{eq:free}
\eea
where $V^{-1}_{S_z S_z^\prime}(\vec k; \vec k^\prime)$ is the
inverse of the interaction matrix defined as
\bea
\sum_{S_z^\prime} \int \frac{d^3k^\prime}{(2\pi)^3}
V_{S_z,S_z^\prime}(\vec k; \vec k^\prime)
V^{-1}_{S_z^\prime,S_z^{\prime\prime}}(\vec k^\prime; \vec k^{\prime\prime})
=\delta_{\vec k, \vec k^{\prime\prime}} \delta_{S_z,S_z^{\prime\prime}}.
\nn
\eea

In order to analyze the pairing eigenvalues,
the gap equation
is linearized around $T_c$ for states around the Fermi surface.
The total angular momentum quantum number $J$ is employed
to classify the eigen-gap functions denoted as
$\phi^{a,JJ_z}_{S_z} (\vec k)$, in which
the index $a$ is used to distinguish different channels with
the same value of $J$.
The eigen-equation for $\phi^{a,JJ_z}_{S_z} (\vec k)$ is
\bea
N_0 \int \frac{d\Omega_{k^\prime}}{4\pi} V_{S_z S_z^\prime} (\vec k; \vec k^\prime)
\phi^{a;JJ_z}_{S_z^\prime} (\vec k^\prime)
=w^a_{J} \phi^{a;JJ_z}_{S_z}(\vec k),
\eea
where %$N_0=\frac{mk_f}{\pi^2\hbar^2}$ is the density of state at
%the  Fermi surface;
$w^a_J$ are dimensionless eigenvalues;
$\vec k,\vec k^\prime$ are at the Fermi surface.
Employing the spherical harmonics decomposition of Eq. \ref{eq:partialwave},
the most negative eigenvalue is calculated lying the channel of $J=L=1$ as
$w^{J=1}=-2\pi\lambda_m$.
All other negative eigenvalues are significantly smaller.
Thus, the dominant
pairing channel remains in the $J$-triplet
channel in the weak coupling theory in agreement with the
real space analysis.

For later convenience, the pairing matrix in the $J=1$ sector is
represented as
\bea
\Delta_{\alpha\beta}^\mu(\vec k)=\frac{\Delta}{2}
\epsilon_{\mu\nu\lambda}(k_\nu \sigma_\lambda - k_\lambda \sigma_\nu)_{\alpha\beta},
\eea
for $\mu=x,y$ and $z$.
It represents a pairing symmetry whose angular momentum projection
along the direction of $\hat e_\mu$ is zero,
{\it i.e.},  it is an eigenstate of $\hat e_\mu \cdot \vec J$
with the zero eigenvalue.

%------------------------------------------------
\subsection{Helical polar pairing and chiral axial pairing}
\label{sect:mgdp_topo}

In the sector of $J=1$, there are still two non-equivalent pairing
possibilities: $J_z=0$, or $J_z=\pm 1$, whose pairing matrices
are $\Delta^z_{\alpha\beta}(\vec k)$ and $\frac{1}{\sqrt 2}
\{\Delta^x_{\alpha\beta}(\vec k) \pm i \Delta^y_{\alpha\beta}(\vec k)\}$,
respectively.
Based on the GL analysis up to the quartic order of
the pairing order parameter, it can be proved that these two are
the only non-equivalent pairing symmetries under 3D rotations.
Right at $T_c$, the Ginzburg-Landau free energy can be linearized,
and these two instabilities are degenerate, while this degeneracy
is lifted below $T_c$ due to the non-linearity of the Ginzburg-Landau
free energy.

In quantum mechanics, if a system possesses rotation symmetry, of course,
%that
the eigenstates of
its energy %level
in the $J=1$ sector and all of their superpositions %suppositions
are degenerate because of %due to
the linearity of quantum mechanics.
However, the index $J$ %to label
labeling the pairing order parameter is not
the angular momentum of the many-body eigenstate.
The description of a many-body system in terms of order parameters
is a great simplification by only keeping a very limit but essential
amount of degrees of freedom.
A price to pay is that the description in terms of
order parameters, say, the Landau-Ginzburg free energy, is non-linear
even though quantum mechanics remains linear.
This is the reason why the $J_z=0$ and $J_z=\pm 1$ in principle
are non-equivalent and the superposition law does not hold
for Cooper pairing symmetries.
The two pairing patterns with $J_z=1$ and $J_z=-1$ are equivalent to
each other which can be connected either by a rotation
or by the TR transformation.

%-------------------------------------------------------------------
\begin{figure}
\centering\epsfig{file=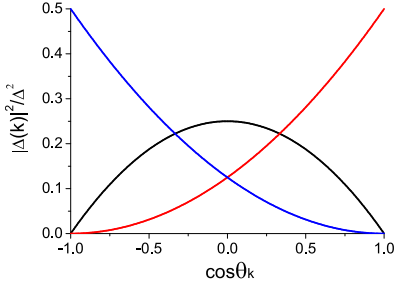,clip=1,width=0.6\linewidth, angle=0}
\caption{The angular distribution of the gap function $|\Delta(\vec k)|^2$
v.s. $\cos\theta_k$ in the helical polar pairing state (the red line)
and the axial pairing state (the black lines).
From Ref. [\onlinecite{li2012}].
}
\label{fig:gap}
\end{figure}
%------------------------------------------------------------------

\subsubsection{Helical polar pairing}
The pairing symmetry with $J_z=0$ is also called the $J$-polar pairing,
characterized by the following pairing matrix
$\Delta^{pl}_{\alpha\beta}=\frac{1}{2}|\Delta|[ k_y \sigma_1 -
k_x \sigma_2) i\sigma_2 ]_{\alpha\beta}$ as
\bea
\Delta_{\alpha\beta}^{pl}&=&
\frac{|\Delta|}{2} \left[
\begin{array}{cc}
\hat k_x -i \hat k_y&0  \\
0& \hat k_x+i \hat k_y
\end{array}
\right],
\eea
which is a unitary pairing and preserves TR symmetry.
The spin-up fermions are paired in
%the way of
the $p_x-ip_y$ symmetry, while, the spin-down fermions
are paired with the $p_x+ip_y$ symmetry. %, and thus, it is equivalent
Thus, it is equivalent to a helical version of the $^3$He-$A$ phase.
In the $^3$He-$A$ phase, the pairing symmetry defines an orbital
angular momentum direction represented by the $l$-vector.
Here, the $z$-axis plays a similar role, but due to the TR
symmetry, it represents the bi-direction of the polar axis, and thus
it is no long a vector but a director.
Thus if we rotation the polar pairing around any axis in the
$xy$-plane for 180$^\circ$, then the system returns to itself
up to a global phase difference.

The Bogoliubov quasiparticle spectra are degenerate for two different spin
configurations as $E^{pl}_{k,\alpha}=\sqrt{\xi^2_k+|\Delta^{pl}(\vec k)|^2}$
with the anisotropic gap function $|\Delta^{pl}(\vec k)|^2=
\frac{1}{4}|\Delta|^2 \sin^2\theta_k$
depicted in Fig. \ref{fig:gap}.
The effective linearized Hamiltonian around the north and south poles
can be combined as
\bea
H_{\uparrow\uparrow} = \psi^\dagger (\vec k)
\left(
\begin{array}{cc}
v_f(k_z-k_f)& k_x-i k_y\\
k_x+i k_y &-v_f(k_z-k_f)
\end{array}
\right) \psi(\vec k), \ \ \,
\label{eq:weyl-pl}
\eea
where $\psi^T(\vec k)=(\psi_\uparrow(\vec k), \psi_\uparrow^\dagger(-\vec k))$.
Eq. \ref{eq:weyl-pl} describes a Weyl fermion with a unit
winding number.
Combining the effective Hamiltonian for the spin down sector
which is also a Weyl fermion but with an opposite chirality,
the low energy Bogoliubov spectrum is 3D gapless Dirac like.
If the quantization axis for the polar pairing is rotated
away from the $z$-axis, then the spin quantization axis for the Bogoliubov
quasiparticles should also be transformed accordingly.

%-------------------------------------------------
\subsubsection{The chiral axial pairing}
Without loss of generality, we pick up the $J_z=1$ pairing, and the
result of the state of $J_z=-1$ can be obtained by performing  TR
transformation.
The pairing matrix,
$\Delta^{ax}_{\alpha\beta}=\frac{1}{2\sqrt 2}|\Delta|
\{ \hat k_z (\sigma_1 + i\sigma_2) i\sigma_2
+ i \sigma_3 \sigma_2  (\hat k_x+i \hat k_y) \}_{\alpha\beta}$,
takes the form of
\bea
\Delta_{\alpha\beta}^{ax}(\vec k)
&=&\frac{\sqrt 2}{2}|\Delta| \left[
\begin{array}{cc}
\hat k_z&\frac{1}{2} (\hat k_x+i \hat k_y)  \\
\frac{1}{2} (\hat k_x+i \hat k_y)& 0
\end{array}
\right], \ \ \, \ \ \,
\eea
%which satisfies
%\bea
%\Delta^\dagger (\vec k)\Delta (\vec k)=
%|\Delta|^2 \left[\frac{1}{2}(1+\hat k_z^2)+\hat k_z (\hat k \cdot
%\vec \sigma)\right],
%\label{eq:axial_pair}
%\eea
thus, this is a non-unitary pairing state.
The Bogoliubov quasiparticle spectra have two non-degenerate branches
with anisotropic dispersion relations as
$E^{ax}_{i}(\vec k)=\sqrt{\xi^2_k+|\Delta^{ax}_i(\vec k)|^2}$,
with
\bea
|\Delta^{ax}_i (\vec k)|^2=
\frac{1}{8} |\Delta|^2 (1\pm \cos\theta_k)^2
\eea
for $i=1$ and $2$, respectively, as depicted in Fig. \ref{fig:gap}.

Considering the coupling between $\vec k$ and $-\vec k$, we can combine
the north and south poles together into a four component spinor.
The above energy spectra show that two of them are gapped, while,
another two
of them are gapless forming a two-component Weyl spinor.
Different from the usual linear dispersion of Weyl fermions, the
dispersions are quadratic with respect to the the transverse
momentum $k_\pp=\sqrt{k^2_x+k^2_y}$.

A natural question is: % that
which pairing is energetically more stable? %.
For the case of $^3$He, at the mean-field level,
the $B$-phase %at the mean-field level
is always
more stable than % the
the $A$-phase \cite{balian1963} because the gap function
of the $B$ phase is %are
uniform over %on
 the Fermi surface.
This can be intuitively understood as follows: the BCS mean-field
free energy density in momentum space ${\cal F}[|\Delta(\vec k)|^2]$
can be viewed as a functional of the gap function $|\Delta(\vec k)|^2$.
Let us consider the pairing on the Fermi surface with the
constraint of $\int d\Omega_k |\Delta(\vec k)|^2$ fixed as a constant,
and then to minimize the free energy.
Usually, the non-linearity of the free energy favors a distribution
of $|\Delta_k|^2$ as uniform as possible.
Here, the situation is quite subtle.
As shown in Fig. \ref{fig:gap}, for both cases, the pairing gap function
distributions on the Fermi surface are non-uniform.
Nevertheless, the distribution of the unitary polar pairing is more
uniform than that of the axial pairing.
Naturally, we expect that, at the mean-field level, the polar pairing
wins.
This has been numerically confirmed by comparing the BCS mean-field
free energies of both pairings in Ref. [\onlinecite{li2010}].
However, we need to bear in mind that this conclusion is only
valid at the mean-field level.
We cannot rule out the possibility that certain strong coupling effects
can stabilize the axial state.
In fact, even in the $^3$He case, the $A$ phase wins in certain
parameter regime in which the strong correlation spin feedback
effect dominates \cite{leggett1975,brinkman1974}.

\subsubsection{More discussions}
The above study of the spin-$\frac{1}{2}$ magnetic dipolar system is just
a toy model to start with.
Of course, the energy scale is too small to be observed in
current cold atom systems.
Even for large moment atoms $^{161}$Dy ($\mu=10\mu_B$)
with the current available density  $10^{13}$ cm$^{-3}$,
the dipolar energy scale is only around 6$nK$ \cite{fregoso2010}.
One possible way to enhance the interaction energy scale is to impose
the optical lattice structure.
For particles in the same lattice site, the inter-particle distance is at the
order of $100nm$, which can enhance the interaction energy scale
to $0.6\mu K$.
Even the $s$-orbital band can hold up to $2F+1$ fermions per site.
We expect that this system can exhibit a variety of even more exotic
many-body physics to be explored.

%%%%%%%%%%%%%%%%%%%%%%%%%%%%%%%%%%%%%%%%%%%%%%%%%%%%%%%%%%%%%%%
%\section{\small A novel spin-dependent dipolar interaction}
%Recently, a new two-component dipolar system was experimentally
%realized in J. Ye's group \cite{yan2013}.
%$|JJ_z\rangle$ are used to denote the rotation eigenstates of the
%RbK molecule.
%The two components $|00\rangle$ and $|11\rangle$ are taken
%as spin $\uparrow$ and $\downarrow$, respectively.
%They are coupled through circularly-polarized microwaves which
%give rises to a spin-dependent dipolar interaction
%\bea
%V(\vec r_1-\vec r_2) = \frac{d^2}{|\vec r_1-\vec r_2|^3}
%P_2(\cos\theta_r) (S_{1,x} S_{2,x}+S_{1,y} S_{2,y}).
%\label{eq:dipolar_2}
%\eea
%We plan to study the pairing symmetry in this system, in particular,
%in the spin-triplet channel.
%The interaction Eq. \ref{eq:dipolar_2} disappears in the channel of
%of total spin $(S=1, S_z=\pm 1)$, and thus the triplet pairing
%can only exist in the channel of  $(S=1, S_z=0)$.
%and the dipolar interactions.

%%%%%%%%%%%%%%%%%%%%%%%%%%%%%%%%%%%%%%%%%%%%%%%%%%%%%%%%%%%%
\section{Conclusions}
\label{sect:conclusion}

We have briefly reviewed the novel many-body physics of both electric
and magnetic dipolar fermion systems focusing on the aspect of
unconventional symmetries.
The electric dipolar interaction is characterized by its
$d_{r^2-3z^2}$-anisotropy, which leads to anisotropic Fermi
liquid properties in the particle-hole channel and a
robust mechanism of spin-triplet $p_z$-wave Cooper pairing
in the particle-particle channel.
The competition and coexistence between the singlet and
triplet Cooper pairing leads to a novel mechanism of
TR symmetry breaking pairing.
The magnetic dipolar interaction manifests its SO
coupled nature in unpolarized fermion systems.
Its Fermi liquid theory is SO coupled whose collective
zero sound mode exhibits an oscillation of a topological
non-trivial spin structure over the Fermi surface.
The magnetic dipolar interaction also lead to a
SO coupled $p$-wave spin triplet Cooper pairing state
with the total spin of a Cooper pair $J=1$.
This is a novel pairing symmetry different from that in both $^3$He-$B$
and $A$ phases, whose Bogoliubov quasiparticles exhibit nodal helical
Dirac or chiral Weyl spectra.

\begin{acknowledgments}
We thank C. K. Chan,  S. Das Sarma, J. Hirsch, W. C. Lee,
and K. Sun for collaborations.
In particular, C. W. is grateful to S. Das Sarma for introducing the field
of dipolar fermions to him.
We also thank E. Fradkin, A. J. Leggett, S. Kivelson, S. C. Zhang,
and F. Zhou for
helpful discussions and encouragements.
C. W. acknowledges the support from AFOSR FA9550-14-1-0168,
NSF-DMR 1410375, and the NSF of China under Grant No. 11328403.
Y. L. thanks the support of the Princeton Center for Theoretical
Science.
\end{acknowledgments}

%%%%%%%%%%%%%%%%%%%%%%%%%%%%%%%%%%%%%%%%%%%%%%%%%%%%%%%%%%%5

%\bibliographystyle{prsty}
%\bibliography{dipolar,dipolar_2,majorana}

\end{document}